\documentclass[fleqn,usenatbib]{mnras}
\usepackage{newtxtext,newtxmath}

\usepackage[T1]{fontenc}
\usepackage{longtable}

\usepackage{graphicx}% Include figure files
\usepackage{bm}% bold math
\usepackage{natbib}
\usepackage{booktabs}
\usepackage{color}
\usepackage{orcidlink}
\usepackage{units}

\usepackage[normalem]{ulem}

\newcommand{\program}[1]{\textsc{#1}}
\newcommand{\citeg}[1]{\citep[e.g.,][]{#1}}
\newcommand\numberthis{\addtocounter{equation}{1}\tag{\theequation}}
\newcommand{\be}{\begin{equation}}
\newcommand{\ee}{\end{equation}}

\title[CSM-interaction lightcurves]{From mass-loss histories to lightcurves: a generalised framework for interaction-powered transients}
\author[N.~Sarin and R.~Hirai]{Nikhil Sarin$^{1, 2}$\thanks{E-mail:nsarin.astro@gmail.com}\orcidlink{0000-0003-2700-1030}, Ryosuke Hirai$^{3,4,5}$\thanks{E-mail:ryosuke.hirai@riken.jp}
\\
$^{1}$Kavli Institute for Cosmology, University of Cambridge, Madingley Road, CB3 0HA, UK\\
$^{2}$Institute of Astronomy, University of Cambridge, Madingley Road, CB3 0HA, UK\\
$^{3}$Astrophysical Big Bang Laboratory (ABBL), RIKEN Pioneering Research Institute (PRI), 2-1 Hirosawa, Wako, Saitama 351-0198, Japan\\
$^{4}$School of Physics and Astronomy, Monash University, Clayton, Victoria 3800, Australia \\ 
$^{5}$OzGrav: The ARC Centre of Excellence for Gravitational Wave Discovery, Australia }

\date{Accepted XXX. Received YYY; in original form ZZZ}

\pubyear{\the\year{}}

\begin{document}
\label{firstpage}
\pagerange{\pageref{firstpage}--\pageref{lastpage}}
\maketitle

\begin{abstract}
We introduce a fast ($\sim 1$--$50$ ms) and generalised public framework for modelling interaction-powered transients. The framework solves the thin-shell equations of motion for ejecta colliding with circumstellar material (CSM), and supports arbitrary CSM density and velocity profiles, including steady winds, eruptions, and complex time-variable mass-loss histories. For optical/UV lightcurves, we implement two luminosity treatments: a fast one-zone mode based on the thin-shell shock power, and a finite-shell transport mode that evolves trapped radiation, photon diffusion, shock emergence, and post-emergence cooling for finite, static CSM shells. In a benchmark comparison, the transport calculation and an optional time-dependent shock-efficiency prescription reproduce the main qualitative and quantitative features of a one-dimensional radiation-hydrodynamical simulation. We use the same shock solution to post-process radio synchrotron and thermal bremsstrahlung X-ray predictions, enabling self-consistent multi-wavelength diagnostics. We show that the assumed CSM velocity structure can significantly affect inferred parameters even
when the density profile at explosion is identical, and that aspherical CSM can mimic multiple spherical shells in bolometric lightcurves. We demonstrate the framework through recovery of a synthetic time-variable mass-loss history and applications to six transients: the Type IIn SN~2010jl, the rapidly evolving stripped-envelope SN~2023xgo, the Type Ia-CSM SN~2020aeuh, the hydrogen-poor superluminous SN~2015bn, the eruptive LBV-like transient SN~2009ip, and the long-duration interacting event iPTF14hls. The inferred CSM structures span steady or enhanced winds, thermonuclear interaction, eruptive density enhancements, and highly structured pre-supernova mass loss, illustrating the framework's utility for inference on upcoming large samples of interacting transients.
\end{abstract}


\section{Introduction}\label{sec:intro}
Observations of core-collapse supernovae provide a glimpse into the final moments of massive stars. Such observations have shed significant light into the core-collapse explosion mechanism~\citep[e.g.,][]{Janka2017}, the conditions in structure of massive stars before core-collapse, and nucleosynthesis in massive stars~\citep[e.g.,][]{Jerkstrand2026}. 

Advances in time-domain instrumentation and larger volume surveys have opened up opportunities to study the explosions of massive stars in ever-increasing detail. In particular, the Vera C. Rubin Observatory is expected to transform our understanding of core-collapse supernovae by providing a large homogeneous sample with deep observations before and many months after the supernova~\citep[e.g.,][]{Gagliano2025}. Such pre-supernova and late-time observations should provide significant clues into another critical aspect of stellar evolution; the mass-loss history of massive stars~\citep[see e.g., for reviews][and references therein]{Smith2014_review, Chandra2018}. 

The mass-loss history of massive stars is intimately connected to their final fate and the observable properties of their core-collapse explosions~\citep{Woosley2018, Owocki2019}. Throughout their evolution, massive stars undergo episodes of enhanced mass loss driven by various physical processes including stellar winds, pulsations, binary interactions~\citep{Podsiadlowski1992,Ercolino2024} and eruptive events such as luminous blue variable (LBV) eruptions or pulsational pair instability~\citep{Smith2014_review}. 
These mass-loss episodes create complex circumstellar medium (CSM) structures that surround the progenitor star at the time of core collapse. The density, composition, and spatial distribution of this CSM encode crucial information about the progenitor's evolutionary history and the physical mechanisms governing mass loss in the final stages of stellar evolution~\citep{Moriya2012, Chevalier2017}.

When supernova ejecta collide with this pre-existing CSM, the resulting shock interaction can power luminous transients that are observable across the electromagnetic spectrum~\citep{Chevalier2017, Tsuna2021}. The luminosity and spectral evolution of these interaction-powered events depend sensitively on the structure and properties of the CSM, making them powerful probes of the progenitor's mass-loss history. Furthermore, the interaction mechanism itself provides further insight into the explosion energy, ejecta mass, and velocity structure of the supernova, offering a complementary diagnostic to traditional supernova observations. The timescales and multi-wavelength signatures of CSM interaction can reveal whether the mass loss occurred as steady winds, discrete eruptions, or complex multi-phase outflows.

Beyond traditional core-collapse supernovae, CSM interaction is responsible for a diverse zoo of transient phenomena. These events span a wide range of luminosities, timescales, and spectral properties, suggesting that CSM interaction operates across different progenitor masses, explosion energies, and mass-loss histories. Understanding this diversity requires a unified theoretical framework that can account for the full parameter space of CSM properties and explosion characteristics while remaining computationally tractable for population synthesis studies and observational inference.

Generally, current models for CSM interaction often rely on simplified assumptions about the CSM structure or are limited to specific regimes of the parameter space~\citep[e.g.,][]{Chatzopoulos2013, Villar2017, Jiang2020, Zhang2025, Sarin2025}.  This helps cast the generic numerical problem into self-similar solutions that can be solved analytically, or without significant numerical computation~\citep{Chevalier1982, Moriya2013}. While these approaches have provided valuable insights, the increasing diversity of observed interaction-powered transients and the prospect of large statistical samples from upcoming surveys demand more flexible and comprehensive modelling frameworks that can also be evaluated relatively quickly, to enable statistical inferences at the anticipated upcoming scale. Such models must be capable of handling arbitrary CSM density profiles, multiple interaction zones, and time-dependent energy injection while remaining sufficiently efficient for systematic studies of large samples. A further growing need is for self-consistent estimates of radio and/or X-ray emission such that observations of facilities such as Einstein Probe~\citep{Yuan2015} and the Square Kilometer Array~\citep{Dewdney2009} can be placed in the correct context.

This paper is aimed at addressing this need for computationally tractable but flexible and self-consistent models for interacting transients. In particular, in Sec.~\ref{sec:model}, we introduce the basic ingredients of a generalized framework for modelling the lightcurve from the interaction of ejecta with pre-existing CSM. We explicitly separate the faster, one-zone mode from the finite-shell radiation-transport mode, and compare these treatments against a hydrodynamical lightcurve to clarify their domains of applicability. In Sec.~\ref{sec:multi}, we discuss the radio and X-ray signatures of interaction-powered transients that can be calculated from the same shock evolution to provide self-consistent, multi-wavelength diagnostics. In Sec.~\ref{sec:inference}, we test this model by inferring properties from different types of interaction-powered transients, which provides a validation test for the various regimes of the general framework. We conclude by discussing the limitations and implications of our model for CSM-interaction modelling in Sec.~\ref{sec:discussion}.
%%%%%%%%%%%%%%%%%%%%%%%%%%%%
\section{Model}\label{sec:model}
We start by considering the general scenario of some outflow (either ejecta launched from a supernova, eruption or strong stellar winds) which collides with pre-existing CSM, produced as a result of some simple (or complex) mass-loss history. If there is efficient radiative cooling, this interaction can produce a thin shell which separates the unshocked CSM (forward shock) and explosion ejecta (reverse shock). Assuming the shell is geometrically thin, the time evolution of the shell is set by momentum conservation \citep{Chevalier1982, Moriya2013},
\begin{align*}
\frac{dM_{\rm sh}}{dt} &= 4\pi R_\mathrm{sh}^2 \rho_{\rm expl}(v_{\rm expl} - v_{\rm sh}) \nonumber + 4\pi R_{\rm sh}^2 \rho_{\rm csm}(v_{\rm sh} - v_{\rm csm}), \numberthis \label{eq:evolution1}\\
 \frac{dR_\mathrm{sh}}{dt}&=v_\mathrm{sh}, \numberthis \label{eq:evolution2}\\
M_{\rm sh} \frac{dv_{\rm sh}}{dt} &= 4\pi R_{\rm sh}^2 \rho_{\rm expl}(v_{\rm expl} - v_{\rm sh})^2 \\
&\quad - 4\pi R_{\rm sh}^2 \rho_{\rm csm}(v_{\rm sh} - v_{\rm csm})^2. \numberthis \label{eq:evolution3} \\
\end{align*}
Here, $M_{\rm sh}$, $R_{\rm sh}$, and $v_{\rm sh}$ are the mass, radius, and velocity of the shell, $\rho$ and $v$ are the density and velocity at the location of the shell, with the subscripts ``csm'' and ``expl'' corresponding to pre-existing CSM or the subsequent explosion, respectively.

The luminosity generated by this interaction is the sum of the energy dissipated by the forward and reverse shocks, which can be estimated as, 
\begin{equation}\label{eq:luminosity}
L_{\rm heat} = \epsilon_{\rm FS} \cdot 2\pi R_{\rm sh}^2 \rho_{\rm csm}(v_{\rm sh} - v_{\rm csm})^3 + \epsilon_{\rm RS} \cdot 2\pi R_{\rm sh}^2 \rho_{\rm expl}(v_{\rm expl} - v_{\rm sh})^{3}.
\end{equation}
Here, $\epsilon_{\rm FS}$ and $\epsilon_{\rm RS}$ dictate the efficiency of the conversion from the kinetic energy into radiation from the forward and reverse shocks, respectively. Numerical simulations show that these efficiencies are time-dependent and are not necessarily the same. In our default inference setup, we absorb this uncertainty into a single nuisance parameter, $\epsilon$, applied to both shocks, and marginalise over a broad range motivated by numerical simulations~\citep[e.g.,][]{Owocki2019, Tsuna2019, Tsuna2023}. In addition to the constant-efficiency prescription, we implement a physically-motivated time-dependent efficiency model. This has two pieces. First, for each shock component we estimate whether the shocked gas can cool locally by comparing the free-free cooling time to the shock flow time~\citep{Tsuna2019}. We compute the immediate post-shock density and pressure from the strong-shock jump conditions,
\begin{equation}
\rho_{\rm sh} = \frac{\gamma_{\rm sh}+1}{\gamma_{\rm sh}-1}\rho_{\rm up}, \qquad
p_{\rm sh} = \frac{2}{\gamma_{\rm sh}+1}\rho_{\rm up}v_{\rm rel}^{2},
\end{equation}
where $\rho_{\rm up}$ is the upstream density and $v_{\rm rel}$ is the shock velocity relative to the upstream material. In the implementation we set $\gamma_{\rm sh}=5/3$ and approximate the free-free volumetric cooling rate as
\begin{equation}
q_{\rm ff}=C_{\rm ff} p_{\rm sh}^{1/2}\rho_{\rm sh}^{3/2},
\end{equation}
where $C_{\rm ff}$ is the cgs normalisation used in the code after absorbing order-unity composition factors. The parameter $\epsilon_{\rm max}$ is the user-supplied maximum shock-to-radiation efficiency. We then define 
\begin{equation}
\epsilon_{\rm cool}(t) =
\epsilon_{\rm max}
\min\left[
1,\,
\frac{q_{\rm ff} R_{\rm sh}}
{3 p_{\rm sh} v_{\rm rel}}
\right].
\end{equation}
This corresponds to $\epsilon_{\rm max}\min(1,t_{\rm flow}/t_{\rm cool})$, with $t_{\rm flow}\simeq R_{\rm sh}/v_{\rm rel}$ and $t_{\rm cool}\simeq 3p_{\rm sh}/q_{\rm ff}$. Thus, shocks that cool rapidly compared to the flow time are allowed to radiate efficiently, while slowly cooling shocks convert only a smaller fraction of their kinetic power into radiation.

Second, once the shock approaches the outer edge of the CSM, the remaining optical depth ahead of the shock becomes small and the newly dissipated energy is less efficiently thermalised into the trapped thermal radiation field. We therefore multiply the local cooling efficiency by a smooth optical-depth thermalisation factor,
\begin{equation}
f_{\tau} = \left[f_{\rm min} + \frac{1-f_{\rm min}} {1 + (\tau_{\rm th}/\tau_{\rm ahead})^{\alpha}} \right] \left[1 - \exp\left(-(\tau_{\rm ahead}/\tau_{\rm thin})^2\right)
\right],
\end{equation}
where $\tau_{\rm ahead}$ is the optical depth from the shock to the outer CSM edge. The first term allows the shock to become only partially radiative as the remaining column decreases, while the second suppresses thermalisation once the upstream column becomes optically thin. In this work we use $f_{\rm min}=0.25$, $\tau_{\rm th}=7$, $\alpha=12$, and $\tau_{\rm thin}=0.1$. We stress that these values are a simple calibration of unresolved radiation-hydrodynamic effects and chosen to create a functional form that best matches the hydrodynamical simulations (see Fig.~\ref{fig:hydro_comparison}) for similar initial conditions, while the transport calculation itself still determines the photon diffusion and escape. 
The time-dependent shock efficiency is then
\begin{equation}
\epsilon_{\rm sh}(t) = \epsilon_{\rm cool}(t) f_{\tau}(\tau_{\rm ahead}).
\end{equation}
This prescription can be evaluated for any CSM profile in the model: it combines a local radiative-cooling condition with the optical-depth dependence of photon thermalisation. However, the detailed functional form remains phenomenological because it absorbs unresolved ionisation, clumping, geometry, frequency-dependent transfer, and multidimensional leakage into a small number of fixed choices. For the inference applications below we therefore use a constant effective efficiency and marginalise over it, rather than imposing this particular time-dependent prescription as a universal, implicit prior.

The thin-shell approximation is most appropriate when the shocked gas cools efficiently enough that the forward- and reverse-shocked regions occupy a small radial width compared with $R_{\rm sh}$. It is less appropriate when shock cooling is inefficient and thus larger fractions of the shock energy is locked up in the shocked region in the form of internal energy instead of being radiated. Alternatively, hydrodynamic and radiative instabilities could broaden the shocked region. In these cases the true forward/reverse shock velocity is above/below the velocity obtained from the thin-shell approximation and could bias the inferred shock velocity and the resulting light curves \citep{Takei2020}.

We first describe the spatially resolved luminosity treatment, which we refer to as the \textit{transport} mode.
In this mode we evolve a radiation-energy-density field on a radial grid spanning the finite CSM domain, i.e. the radial interval over which the CSM density is non-zero, similar to the implementation recently discussed in~\citet{Zhang2025} for \texttt{TransFit-CSM}. The current transport implementation is intended for finite, static CSM shells with well-defined inner and outer edges.
During the interaction phase, the radiation field is sourced at the shell and transported through the material ahead of the shock. We evolve 
\begin{equation}
\frac{\partial e}{\partial t}
=
\frac{1}{r^2}\frac{\partial}{\partial r}
\left(
\frac{c\,r^2}{3\kappa \rho(r,t)}
\frac{\partial e}{\partial r}
\right),
\qquad
R_{\rm sh}(t) < r < R_{\rm out},
\label{eq:diffusion_pde}
\end{equation}
where $R_{\rm out}$ is the outer edge of the CSM.
The inner boundary condition is set by the shock heating rate,
\begin{equation}
-\frac{c}{3\kappa \rho}\frac{\partial e}{\partial r}\Big|_{R_{\rm sh}}
=
\frac{L_{\rm heat}}{4\pi R_{\rm sh}^2},
\label{eq:inner_bc}
\end{equation}
while the emergent luminosity is obtained from the outward diffusive flux at the photosphere,
\begin{equation}
L_{\rm obs}(t)
=
4\pi R_{\rm ph}^2
\left(
-\frac{c}{3\kappa \rho}\frac{\partial e}{\partial r}
\right)_{R_{\rm ph}}.
\label{eq:l_obs_transport}
\end{equation}
When the shock reaches the outer edge of the CSM at $t=t_{\rm se}$, shock heating is switched off. The remaining luminosity is then powered by the radiation energy stored in the shocked material. We follow this phase by allowing the shocked shell to expand homologously with a velocity set by the shell velocity at emergence, while continuing to diffuse the trapped radiation through the expanding material. This cooling phase is less uniquely specified by the thin-shell dynamics than the interaction phase, because the thin-shell approximation does not resolve the internal structure of the shocked shell. We therefore prioritise global energy conservation across shock emergence and evolve the post-emergence radiation reservoir with the same grey diffusion approximation. This treatment is approximate, as we still assume spherical symmetry, grey opacity, and a geometrically thin interaction shell, but it captures photon trapping and delayed release in a self-consistent way. In particular, it naturally produces an early dark phase when the shock is deeply embedded, a diffusion-mediated rise to peak, and the approach to $L_{\rm obs}\approx L_{\rm heat}$ as the CSM becomes optically thin as seen in hydrodynamical simulations and more sophisticated full diffusion solutions~\citep{Zhang2025}.

A computationally cheaper approximation is obtained by reducing the spatially resolved radiation field to a single trapped-energy reservoir. We refer to this as the \textit{one-zone mode}. In this approximation the spatial distribution of radiation energy is not evolved; instead, the CSM enters through a single time-dependent diffusion time. The model is therefore not limited to optically thin CSM, but it loses information about where the radiation is stored within the CSM. Without a provided opacity, or in the limit $t_{\rm diff}\rightarrow0$, $L_{\rm obs}(t)=L_{\rm heat}(t)$ up to the choice of shock efficiencies in Eq.~(\ref{eq:luminosity}). Specifically, the one-zone model evolves
\begin{equation}
\frac{dE}{dt}=L_{\rm heat}(t)-\frac{E}{t_{\rm diff}(t)},\qquad
L_{\rm obs}(t)=\frac{E(t)}{t_{\rm diff}(t)},
\end{equation}
where $E$ is the total trapped radiation energy. We take $t_\mathrm{diff}(t)=\tau(R_\mathrm{sh})R_\mathrm{sh}/c$, with the optical depth evaluated from the shell location through the CSM,
\begin{equation}
 \tau(r)=\int^\infty_r \kappa\rho_\mathrm{csm}(r')dr',
\end{equation}
where $\kappa$ is the opacity. The convolution form used in the code is the corresponding Green's-function solution,
\begin{equation}
    L_\mathrm{obs}(t)=\int_0^t \frac{L_\mathrm{heat}(t')}{t_\mathrm{diff}(t')}\exp{\left[-\frac{t-t'}{t_\mathrm{diff}(t')}\right]}dt',\label{eq:l_obs}
\end{equation}
based on $L_\mathrm{heat}(t)$ computed from the shell evolution calculations Eqs.~(\ref{eq:evolution1})--(\ref{eq:luminosity}).
This preserves the speed and robustness of the thin-shell calculation and can be applied to arbitrary CSM histories, including effectively extended, expanding, or phenomenological CSM structures. Its limitation is that all spatial information about the radiation reservoir is compressed into $t_{\rm diff}(t)$; it therefore cannot capture a true dark phase, the detailed diffusion-mediated peak morphology, or post-emergence cooling of a finite shell. We also ignore adiabatic losses and light-travel-time effects in this one-zone closure. For arbitrary time-dependent CSM we therefore use the one-zone mode, because in those cases the appropriate radiation grid and post-shock cooling geometry are not uniquely specified by the thin-shell dynamics alone.

To evaluate the model outputs, we numerically evolve the system using Eqs.~(\ref{eq:evolution1})--(\ref{eq:evolution3}) for a given $\rho_{\rm csm}$ and $\rho_{\rm expl}$. As our ultimate aim is to produce a computationally-efficient forward model, we use adaptive time-stepping in \program{Fortran}, compute the shock heating using Eq.~(\ref{eq:luminosity}), and optionally evolve the transport calculation described above. 
The public interface exposes this through two explicit luminosity branches. The default \texttt{simple} branch returns the shock-powered luminosity from the thin-shell calculation, and corresponds to the one-zone mode when an opacity is supplied. The \texttt{transport} branch instead uses the radiation solver for the observed bolometric luminosity and stores the shock luminosity separately. 
We also compute useful shell properties such as the shell mass, shell velocity, shell radius, photospheric radius, trapped radiation energy, and optical-depth diagnostics. 
We create a \program{python} interface for the \program{Fortran} module that is available via \program{Redback}~\citep{Sarin2024}\footnote{\program{Redback} is available on GitHub at \url{https://github.com/nikhil-sarin/redback}} to provide a simpler interface for generating lightcurves, calculating other quantities, such as band pass magnitudes, and facilitating inference. 
This flexibility in density and velocity profiles is one of the core differences between our methodology and those that exist before, which are almost exclusive to producing lightcurves for static power-law density CSM profiles~\citeg{Chatzopoulos2013}.

For rapid inference applications where the CSM can be represented by a static finite power-law profile and the explosion by a broken power-law ejecta profile, we also implement a \program{JAX} version of the same calculation as the Fortran code. This is not a replacement for the full \program{Fortran} backend; instead, the \program{JAX} implementation is intended as a fast, differentiable backend for high-throughput inference on GPUs. In both implementations, the same physical setup is used, so differences between the two are driven by JAX/GPU implementation details rather than by changes in the model assumptions. For single lightcurve calls the \program{Fortran} implementation is fast, typically $\sim1$--$50$ ms depending on the luminosity branch and requested outputs. The \program{JAX} backend carries an initial compilation cost such that it is generally up to twice as slow for single evaluations, but after compilation is faster for large batches of static power-law models, especially on GPUs by up to an order of magnitude. It is therefore most useful for batched likelihood evaluations or gradient-based workflows rather than as a more general replacement for the Fortran solver.

Given the generality of the above equations, we can capture complex mass-loss histories by creating different $\rho_{\rm csm}$. For example, the CSM density profile could be a product of different periods of wind-driven mass-loss, a product of multiple eruptions or a mixture. This generality enables us to model a range of transients from e.g., fast blue optical transients that are potentially powered by interaction with dense CSM shells with sharp truncated edges~\citep{Margalit2022}, the Great Eruption of $\eta$ Carinae~\citeg{Frew2004,Smith2006, Smith2008}, to supernovae of different classes with interaction signatures~\citeg{Chevalier2017}, supernova impostors~\citeg{Pastorello2013} and precursors~\citeg{Matsumoto2022}, or luminous red novae~\citeg{Matsumoto2022_novae}. 
We highlight that one of the benefits of our generalised approach is that we can also phenomenologically model mass-loss histories, enabling us to fit observations and directly produce posteriors on the mass-loss history. 
At the same time, we can also take real mass-loss histories from full numerical simulations of mass loss in massive stars~\citep[e.g.,][]{Takei2022} as an input and produce end-to-end results that could then instead provide constraints directly on properties of the progenitor star. 

\subsection{Density and velocity structures}
The luminosity prescriptions above specify how shock power is converted into escaping radiation. The remaining inputs to the thin-shell calculation are the density and velocity fields of the two colliding outflows. At each radius the dynamics only require $\rho_{\rm expl}$, $v_{\rm expl}$, $\rho_{\rm csm}$, and $v_{\rm csm}$ evaluated at the shell position. This makes the framework agnostic about whether the material was produced by a steady wind, a time-variable wind, a discrete eruption, an arbitrary reconstructed density profile, or the time ordering between an eruption and stellar winds.

It is commonly assumed in many studies on interaction-powered supernovae that the CSM has a uniform velocity $v=\mathrm{const.}$, resembling that of a stellar wind for which $v_{\rm csm}=v_{\rm w}$ and the density distribution is,
\begin{equation}
    \rho(r)=\frac{\dot{M}}{4\pi r^2v_\mathrm{w}},\label{eq:wind_density}
\end{equation}
where $r$ is the radius and $\dot{M}$ is the mass-loss rate. The resulting distribution is a simple power-law in radius $\rho\propto r^{-2}$, although there could be more complex structures if the mass-loss rate was time varying. Recently, it is becoming increasingly evident that at least in some supernovae, the CSM was created through eruptive mass loss where the ejecta follow a homologous expansion $v=r/t$, where $t$ is the time since the eruption. The supernova itself also usually follows homologous expansion until it enters the Sedov-Taylor phase. For such explosion-like outflows, the density distribution is determined by the details of the explosion, such as the progenitor structure and explosion energy. For simplicity, the density distribution of these explosion outflows have often been expressed as analytical functions, such as an exponential density distribution,
\begin{equation}
\rho_{\rm expl}(v,t) =  \frac{M_{\rm expl}}{8\pi (v_\mathrm{expl}t)^3 }e^{-v/v_{\rm expl}},
\end{equation}
where $v_{\rm expl} = \sqrt{E_{\rm expl}/(6M_{\rm expl})}$ to mimic eruptive mass-loss \citep{Owocki2019} or a more classic broken power-law profile~\citep{Chevalier1994, Matzner1999}, 
\begin{equation}\label{eq:bpl}
\rho_{\rm bpl}(v,t) = A t^{-3}\begin{cases}
(v/v_*)^{-\delta} & v < v_*, \\
(v/v_*)^{-n} & v \geqslant v_*,
\end{cases}
\end{equation}
where 
\begin{equation}
v_* = \sqrt{\frac{2(5-\delta)(n-5)E_{\rm expl}}{(3-\delta)(n-3)M_{\rm expl}}}, 
\end{equation}
and $\delta$ and $n$, typically have values between $0-2$ and $7-12$, respectively based on different progenitors~\citep{Matzner1999} for a supernova explosion but could have different profiles for different eruption mechanisms, while $A$ is implicitly set such that the integration of the explosion density profile results in a total mass, $M_{\rm SN}$.  Throughout this work, we use the subscript ``SN'' to describe the supernova explosion parameters, while the subscript ``expl'' is for any explosion, with ``bpl'' referring to an explosion with a broken power-law profile and ``expo'' referring to an explosion with an exponential profile.

We note that the framework is agnostic to the ordering of the explosion. In other words, while we have generally only described the case here where an ``explosion'' goes into pre-existing CSM above, we can instead describe this last phase with a wind-like or time-variable outflow rather than only by a homologous supernova ejecta profile. The public implementation of the module includes models for several such scenarios. 

\subsection{Static finite CSM: the transport regime}
\label{sec:static_transport}
The two luminosity treatments above have different domains of applicability. The one-zone mode is the robust and fast option for arbitrary CSM histories, including effectively extended winds, multiple phenomenological shells, and time-dependent or aspherical configurations. The transport mode is instead most useful when the CSM can be represented as a finite, optically thick, static shell, where photon trapping, shock emergence, and the stored cooling reservoir are central to the lightcurve morphology.

As a controlled benchmark for the transport calculation, we consider the classic case of a static finite power-law CSM shell interacting with broken power-law supernova ejecta, which is the common parameterisation adopted by numerical lightcurve models~\citep[e.g.,][]{Chatzopoulos2013, Zhang2025}. This setup is simpler than the general CSM structures we will consider later. However, it serves as an important benchmark and helps isolates the effects of optical depth, diffusion, shock emergence, and post-emergence cooling.

The CSM density is \begin{equation}
\rho_{\rm csm}(r) = \rho_{\rm in}
\left(\frac{r}{R_{\rm in}}\right)^{-s},
\qquad
R_{\rm in} \leq r \leq R_{\rm out},
\end{equation}
with zero density outside this interval. The normalisation $\rho_{\rm in}$ is set by requiring the shell mass to be $M_{\rm csm}$. We use $R_{\rm in}=5\times10^2\,{\rm R_\odot}$ and compare two outer radii: a compact shell with $R_{\rm out}=5\times10^3\,{\rm R_\odot}$ and an extended shell with $R_{\rm out}=5\times10^4\,{\rm R_\odot}$ for a fixed CSM mass of $1\,{\rm M_\odot}$. For each case we vary the CSM density slope over $s=0, 0.5, 1, 1.5,$ and $2$.

The supernova ejecta are described by the broken power-law profile of Eq.~\ref{eq:bpl}, with $M_{\rm ej}=5\,{\rm M_\odot}$, $E_{\rm SN}=10^{51}\,{\rm erg}$, inner slope $\delta=1$, and outer slope $n=10$. We adopt a constant shock-conversion efficiency $\epsilon=1$, opacity $\kappa=0.2\,{\rm cm^2\,g^{-1}}$, and run the calculation in \texttt{transport} mode with 40 radiation zones. 

\begin{figure*}
\includegraphics[width=\textwidth]{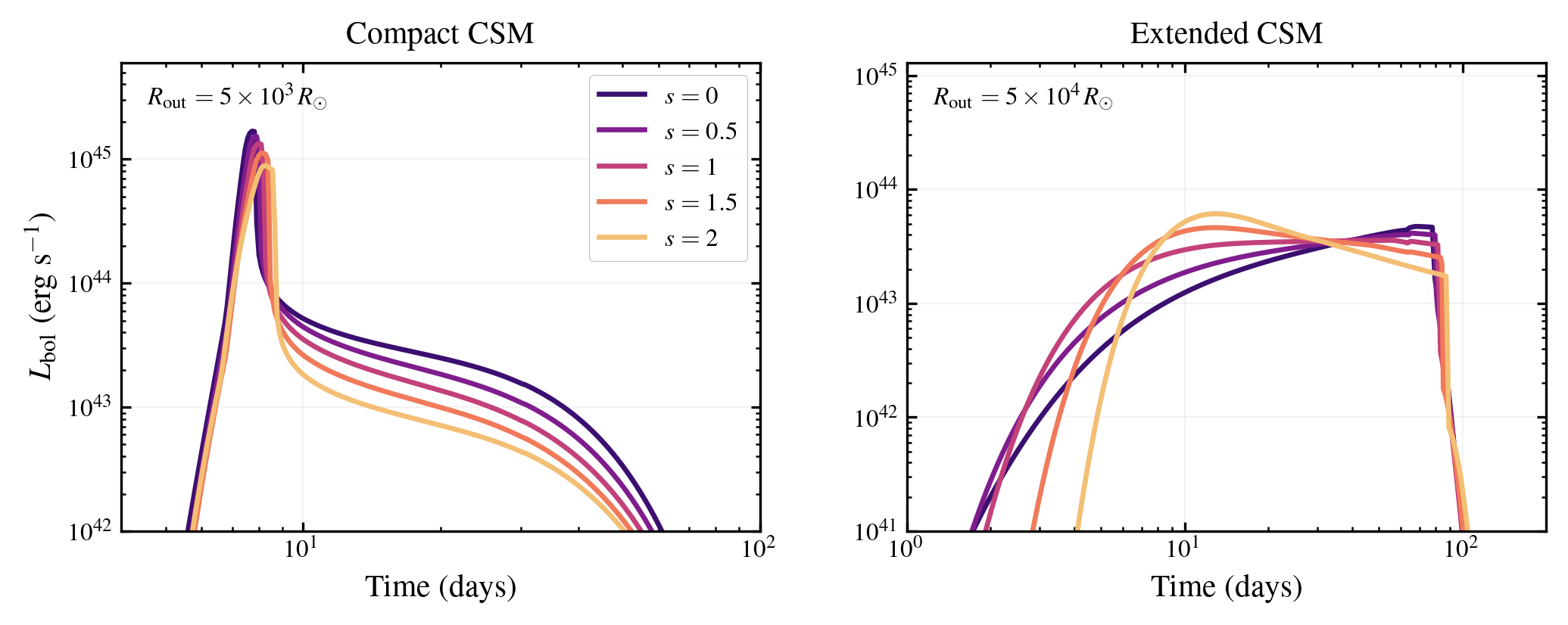}
\caption{Static finite power-law CSM benchmark for the transport solver. The CSM is a finite shell with $M_{\rm csm}=1\,{\rm M_\odot}$ and $R_{\rm in}=5\times10^2\,{\rm R_\odot}$. The left panel shows a
compact shell with $R_{\rm out}=5\times10^3\,{\rm R_\odot}$, while the right panel shows an extended shell with $R_{\rm out}=5\times10^4\,{\rm R_\odot}$. In both panels the CSM density follows $\rho_{\rm
csm}\propto r^{-s}$, with $s=0,0.5,1,1.5,$ and $2$. The ejecta are a broken power-law supernova model with $M_{\rm ej}=5\,{\rm M_\odot}$, $E_{\rm SN}=10^{51}\,{\rm erg}$, $\delta=1$, and $n=10$. The
calculation uses \texttt{transport} mode with $\kappa=0.2\,{\rm cm^2\,g^{-1}}$, constant efficiency $\epsilon=1$, and 40 radiation zones.}
\label{fig:static_transport_benchmark}
\end{figure*}

Figure~\ref{fig:static_transport_benchmark} illustrates the qualitative behaviour expected from the transport mode. In the compact case, shock emergence occurs early and the subsequent luminosity is sensitive
to how much radiation energy remains trapped in the shocked shell. In the extended case, the larger optical depth and longer radial scale produce a more pronounced diffusion-mediated rise. The dependence on
$s$ reflects the fact that changing the density slope changes both the shock dynamics and the location of the radiation reservoir. 
% This benchmark is therefore useful as a numerical diagnostic: changes to the transport solver should preserve the broad ordering, peak times, and cooling morphology shown here.

\begin{figure}
\includegraphics[width=\columnwidth]{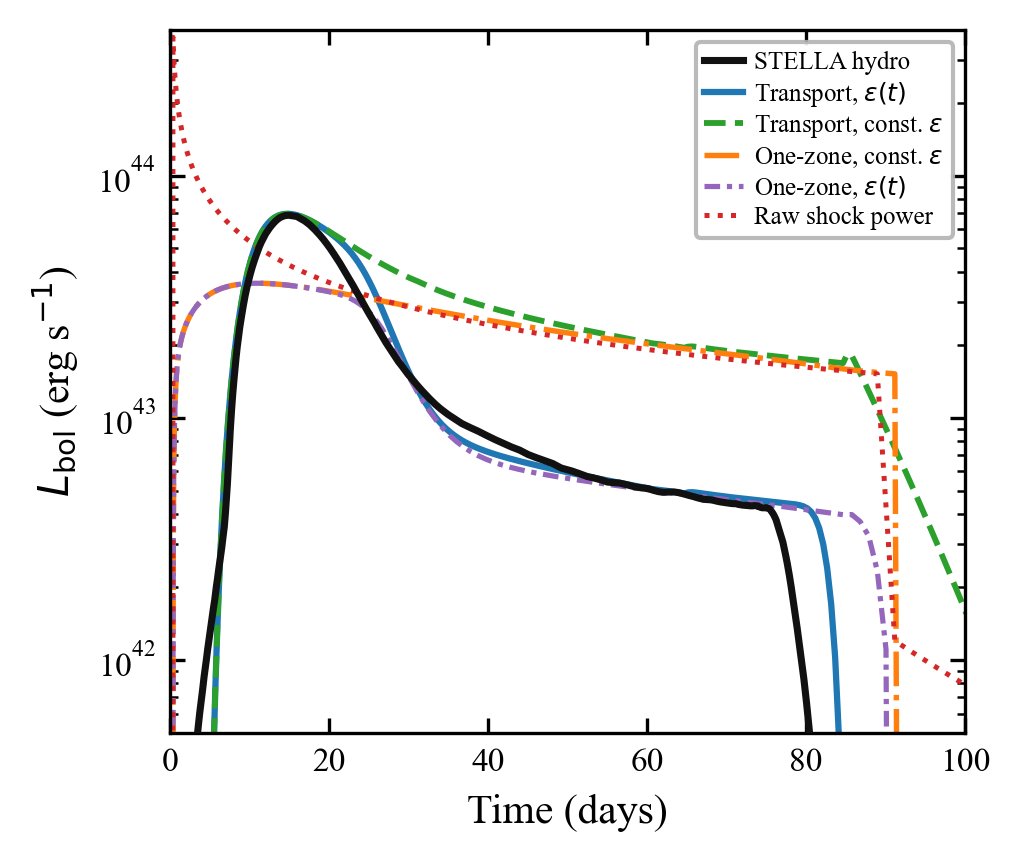}
\caption{Comparison between our semi-analytic luminosity treatments and a one-dimensional radiation-hydrodynamical lightcurve for the same compact CSM configuration. The red curve shows the raw thin-shell shock power before photon diffusion or efficiency losses. The orange and purple curves show the one-zone luminosity with constant $\epsilon=1$ and with the time-dependent $\epsilon(t)$ prescription, respectively. The green and blue curves show the corresponding spatially resolved transport calculations. The hydrodynamical curve is produced with \texttt{STELLA}~\citetext{private communication, Moriya, 2026}; it does not use the same constant grey opacity as our semi-analytic calculations, so part of the residual mismatch is expected from the different radiation-transfer treatment.}
\label{fig:hydro_comparison}
\end{figure}

Figure~\ref{fig:hydro_comparison} provides a more demanding comparison against a one-dimensional radiation-hydrodynamical lightcurve produced with \texttt{STELLA}~\citetext{private communication, Moriya, 2026}. The hydrodynamical calculation can resolve effects that are intentionally compressed into a thin shell in our model and also uses a more detailed radiation-transfer treatment than our grey-opacity calculation, so exact agreement in the detailed lightcurve shape is not expected. However, the comparison is nevertheless diagnostic. The one-zone calculation can reproduce the broad shock-powered decline once the shock is visible, and adding $\epsilon(t)$ largely accounts for the late-time fading by reducing the fraction of newly dissipated power that is thermalised into thermal radiation as the optical depth ahead of the shock falls. The one-zone calculation cannot reproduce the early dark phase or the diffusion-mediated rise to peak; these require following the spatial distribution of the radiation energy, as is done in transport mode. When combined with the time-dependent efficiency prescription, the transport calculation is in much closer qualitative agreement with the hydrodynamical simulation.

The above comparisons provide some practical validation for the one-zone and transport modes. The transport calculation captures photon trapping and the peak morphology, while the time-dependent efficiency mainly controls the late-time thermalisation of shock power and consistency with hydrodynamical simulations after peak. This would suggest that the best path is to use the transport solver with a time-dependent efficiency at all times. However, this method is restricted to finite, static CSM and is up to $\sim 5$ times slower than the equivalent one-zone output. The faster one-zone mode does not capture the peak luminosity or the dark phase but recovers the broad luminosity scale and timescale of the hydrodynamical calculation while remaining flexible and fast enough for high-dimensional inference with arbitrary CSM profiles. For real-transient inference, we therefore advocate comparing results across modelling assumptions, identifying which posterior features are robust, or explicitly marginalising over uncertain choices such as the shock efficiency.

Having defined the finite-shell transport regime above, we now turn to the broader purpose of the framework: mapping flexible mass-loss histories into interaction lightcurves. The examples below generally use the one-zone mode. These examples include effectively extended winds, homologously expanding eruption snapshots, multiple phenomenological shells, and aspherical decompositions. Forcing the finite-shell transport solver onto all of these cases would introduce additional assumptions about the outer boundary, the post-emergence cooling reservoir, and multidimensional diffusion that are not part of the mass-loss model itself.

\begin{figure*}
\includegraphics[width=\textwidth]{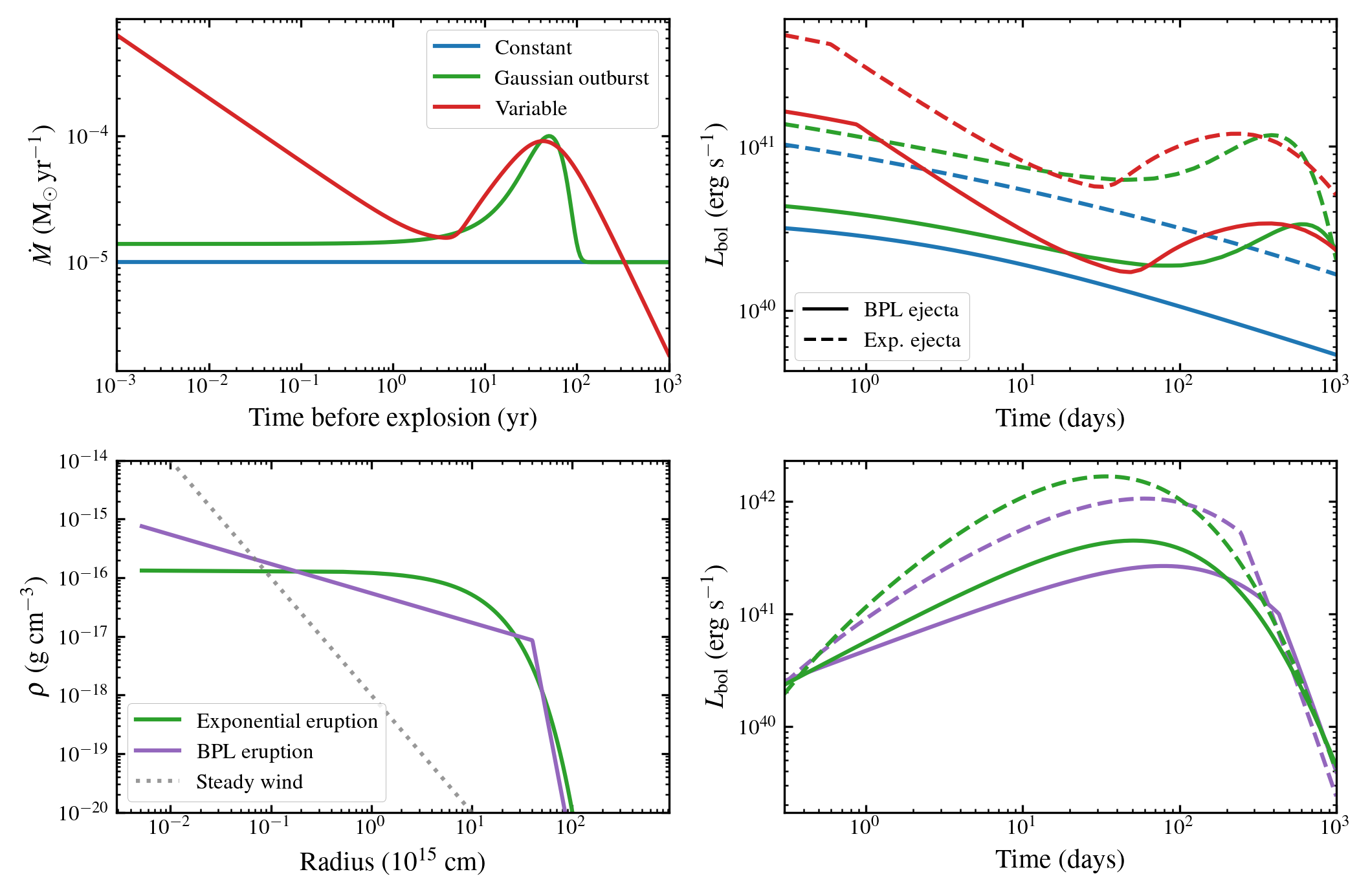}
\caption{Overview of CSM interaction model capabilities demonstrating the diversity of observable signatures from different mass-loss scenarios. \textbf{Top left:} Three representative mass-loss histories: a constant stellar wind at $\dot{M}=10^{-5}\,{\rm M_\odot\,yr^{-1}}$, a Gaussian outburst peaking 50 yr before explosion with $\dot{M}_{\rm peak}=10^{-4}\,{\rm M_\odot\,yr^{-1}}$, and a smooth triple power-law mass-loss history with breaks at 5 and 50 yr. \textbf{Top right:} Corresponding interaction lightcurves when supernova ejecta collide with these CSM structures. Solid lines show broken power-law supernova ejecta ($\delta=0.5$, $n=12$), while dashed lines show exponential ejecta profiles. \textbf{Bottom left:} Density profiles for different eruption types that can create CSM, including exponential and broken power-law eruption profiles, alongside a typical $\rho\propto r^{-2}$ stellar wind for reference. \textbf{Bottom right:} Lightcurves for eruption-eruption interactions, where an earlier eruption with $M_{\rm eruption}=0.05\,{\rm M_\odot}$ and $E_{\rm eruption}=0.01$ foe creates CSM that is subsequently impacted by a more energetic supernova. Purple and green curves show broken power-law and exponential CSM profiles, respectively; solid and dashed lines distinguish BPL and exponential supernova ejecta profiles as in the top-right panel. Unless stated otherwise, the overview examples use a $2\,{\rm M_\odot}$, 1 foe supernova, one-zone luminosities with $\kappa=0.1\,{\rm cm^2\,g^{-1}}$, and $\epsilon=0.4$.}
\label{fig:model_overview}
\end{figure*}

In the top panels of Fig.~\ref{fig:model_overview}, we show the broad capabilities and flexibility of our generalised framework by generating lightcurves for a diverse range of wind mass-loss histories followed by an exponential (dashed) or broken power-law density profile explosion (solid). In particular, here we parameterise the mass-loss history with either a constant mass-loss rate at a specific velocity (blue curves), or a mass-loss rate with a period of increased mass-loss (as a Gaussian) in green, or with a variable mass loss history parameterised by three smooth power-laws in red. For all curves, we assume the same properties of the final eruption; a supernova with $M_{\rm ej} = 2M_{\odot}$ and an explosion energy of $E_{\rm SN} = 10^{51}$erg. The corresponding lightcurves highlight how increases in mass-loss rate translate into features in the lightcurve (with timescales set largely by the wind and supernova velocities), while the interaction with the different supernova density profiles shapes the early lightcurve. Already, we see that diversity in mass-loss histories for the same supernova explosion can explain significant differences in the lightcurve at a range of timescales, illustrating the role of mass-loss history in shaping supernova lightcurves. 
We do emphasise that here for the sake of simplicity we are not showing energy produced by radioactive decay. For the fiducial $2\,M_\odot$, 1 foe ejecta used in Fig.~\ref{fig:model_overview}, a diffusion-processed radioactive component with $M_{\rm Ni}\sim0.1\,M_\odot$ peaks at $\sim10^{42}$--$2\times10^{42}$ erg s$^{-1}$ for typical stripped-envelope opacities. This would exceed the low-density wind-interaction examples and be comparable to the brightest eruption-interaction examples, while a thermonuclear-like yield of $M_{\rm Ni}\sim1\,M_\odot$ would be an order of magnitude brighter still. The relative importance of interaction and radioactivity is therefore strongly event dependent, and in the real-object fits below we include radioactive heating when it is physically motivated.

In the bottom panels of Fig.~\ref{fig:model_overview}, we show density profile snapshots and lightcurves for eruptive mass-loss followed by an eruption. Hereafter we refer to eruptive mass loss as freely expanding outflows with $v=r/t$ velocity profiles. 
We also show a constant wind mass-loss density distribution for comparison purposes. In general, we see that these eruptive mass-loss history lightcurves are brighter than the wind mass-loss scenario, this is a by-product of their larger density compared to the wind mass-loss case, and the fact that the luminosity is directly proportional to this quantity. 

The bottom panels also showcase the rich parameter space accessible through eruption-eruption interactions, where earlier mass-loss episodes create structured CSM that subsequently interacts with the final supernova explosion. The density profile of the CSM-creating eruption fundamentally determines the interaction physics: exponential profiles ($\rho \propto e^{-v/v_0}$) typically produce more gradual evolution as the shock encounters smoothly declining density, while broken power-law profiles with their characteristic velocity breaks can create sharper temporal features when the shock transitions between the steep inner ($\rho \propto v^{-\delta}$) and shallow outer ($\rho \propto v^{-n}$) density regimes.

The supernova ejecta profile itself plays an equally important role in determining the observable signatures. Exponential ejecta (dashed lines) generally produce more extended interaction phases due to their extended velocity distribution, while broken power-law profiles (solid lines) can create more rapid lightcurve evolution, particularly for steep outer slopes ($n \gtrsim 10$). The bottom-right panel shows a more modest but still important spread, with peak luminosities differing by about an order of magnitude and peak times by a factor of a few for the parameter choices shown. This highlights how small changes in the outflow density profile alone can noticeably alter both the luminosity scale and temporal morphology for otherwise similar parameters.

This diversity of signatures seen across Fig.~\ref{fig:model_overview} highlights both the diagnostic power and the interpretational challenges inherent in interaction-powered transients. While the rich phenomenology provides multiple observational handles for constraining the progenitor properties and mass-loss history, the high-dimensional parameter space requires robust modelling frameworks capable of exploring the full range of physically motivated scenarios. It also stresses the need to connect inferred mass-loss and eruption histories with more detailed simulations or theoretical expectations. 

\begin{figure*}
\includegraphics[width=\textwidth]{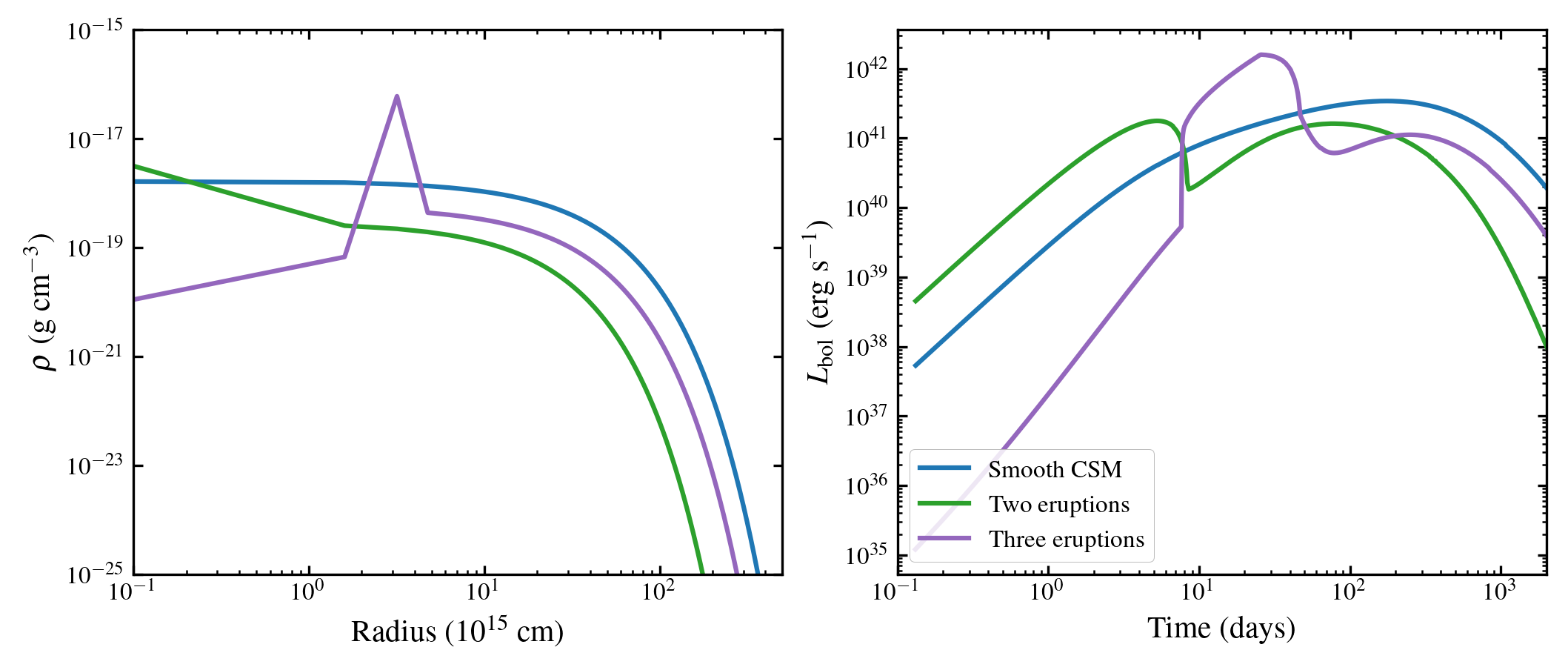}
\caption{Multi-eruption CSM scenarios demonstrating the diversity of interaction signatures from complex pre-supernova mass-loss histories. \textbf{Left:} Circumstellar density profiles for three representative scenarios: a single smooth eruption (blue), two eruptions producing one additional dense shell (green), and a three-eruption sequence producing multiple density enhancements (purple). The blue model uses a single exponential outflow with $M_{\rm csm}=0.22\,{\rm M_\odot}$ and $E_{\rm csm}=0.05$ foe. The green model combines a low-mass exponential component ($0.006\,{\rm M_\odot}$, $4\times10^{-4}$ foe) with a denser $0.035\,{\rm M_\odot}$ shell centred at $0.8\times10^{15}$ cm. The purple model contains an initial $0.04\,{\rm M_\odot}$ exponential outflow plus two additional shells at $3\times10^{15}$ cm and $4\times10^{14}$ cm. \textbf{Right:} Corresponding one-zone bolometric lightcurves for a $2\,{\rm M_\odot}$, 1 foe broken power-law supernova with $\delta=0.5$, $n=12$, $\kappa=0.1\,{\rm cm^2\,g^{-1}}$, and $\epsilon=0.4$. The CSM snapshots use $t_{\rm int}=1$ yr unless otherwise specified, so the homologous velocity associated with a density feature is set by $v=r/t_{\rm int}$ at explosion. The smooth CSM produces a single broad interaction episode, while the multi-eruption cases show how density enhancements can generate sharper bumps or shoulders in the lightcurve.}
\label{fig:multi_eruption_scenarios}
\end{figure*}

\subsection{Modelling multiple eruptions}
We now turn our attention to scenarios where the CSM is a product of multiple eruptions, which is vital for modelling the lightcurves of pulsational pair instability supernovae or core-collapse of massive stars with extensive eruptive mass-loss episodes. Multiple dense shells are expected to form at each interface between the multiple eruption ejecta, which can create distinct bumps in the light curve as it interacts with the final explosion ejecta.

Here, we make some additional simplifying assumptions to keep our forward model as computationally efficient as possible. In particular, we first assume that the gap between the eruptions is shorter than the gap between the final explosion (e.g., supernova) and the preceding eruption. 
This implies that we can assume the CSM is in a steady state by the time the final explosion itself occurs and thus can be characterized with a single $v=r/t$ velocity profile. 
Second, while in reality the asymptotic location of interaction shells can be computed based on momentum balance between the eruption density profiles, we choose to keep the shell location and width as free parameters. This way, we can keep the model flexible enough to also represent non-asymptotic density profiles. 
When creating an arbitrary CSM density profile, we assume (by default) that the provided density profile (density as a function of radius) is a snapshot of the CSM at the time of the supernova ($t=0$). This snapshot is linked to the progenitor's mass-loss history through the interval parameter, $t_{\rm int}$. This interval represents the age of the CSM, interpreted as the time between the supernova and the final mass-loss event that shaped the profile. Consequently, $t_{\rm int}$ implicitly defines the CSM's velocity structure, where a feature at radius $r$ has an expansion velocity of $v = r / (t+t_{\rm int})$.

Fig.~\ref{fig:multi_eruption_scenarios} demonstrates the rich diversity of interaction signatures achievable through multi-eruption scenarios. As with Fig.~\ref{fig:model_overview}, we show the density profile on the left and the corresponding lightcurve on the right. The three cases shown are deliberately simple enough to read visually, but still span smooth CSM, a single additional dense shell, and a more complex multi-shell structure.

The resulting lightcurves showcase fundamentally different temporal evolution depending on the shell locations and CSM structure. The single smooth eruption produces a broad lightcurve, while the two- and three-eruption cases show additional structure as the shock encounters local density enhancements. This illustrates a useful feature of the framework: once the CSM is specified as a density snapshot, the same thin-shell calculation maps the radial structure into a time-dependent shock luminosity. It also illustrates a limitation. Even this reduced set of examples is already high dimensional, since each extra shell requires at least a location, width, density or mass scale, and a velocity or age interpretation, in addition to the supernova parameters. As we will discuss later, fully exploring this parameter space for real data requires either physically motivated priors or more powerful sampling frameworks.

\subsection{Importance of velocity profile}

A common assumption in many interaction-powered SN modelling studies is that the CSM can be described as a steady wind with a constant velocity. Under this assumption, the CSM density follows a $\rho\propto r^{-2}$ (Eq.~\ref{eq:wind_density}) profile, although more complex structures can be represented by allowing for time-dependent mass-loss rates, $\dot{M}(t)$. For this reason, the term ``mass-loss rate history'' is often used interchangeably with the CSM density profile, implicitly assuming a constant CSM velocity. In that case, the density normalisation entering Eq.~(\ref{eq:evolution1}) and Eq.~(\ref{eq:luminosity}) is set by the ratio $\dot{M}/v_{\rm wind}$, while the relative shock velocity is $v_{\rm sh}-v_{\rm wind}$. 

However, in some SNe it is clear that the CSM was produced by eruptive events rather than steady winds. In such cases, the ejecta are expected to expand homologously, with a velocity profile $v=r/t$. A prominent example is SN2006jc, where a luminous eruption was observed approximately two years prior to the SN explosion that exhibited clear signatures of interaction \citep{Foley2007}. For such systems, assuming a constant-velocity, wind-like CSM can be inappropriate. A homologously expanding CSM with the same instantaneous $\rho(r)$ has a velocity distribution $v=r/t$ instead of $v=v_\mathrm{wind}$, so the shock power can differ even if the density snapshot is identical.

\begin{figure}
\includegraphics[width=\columnwidth]{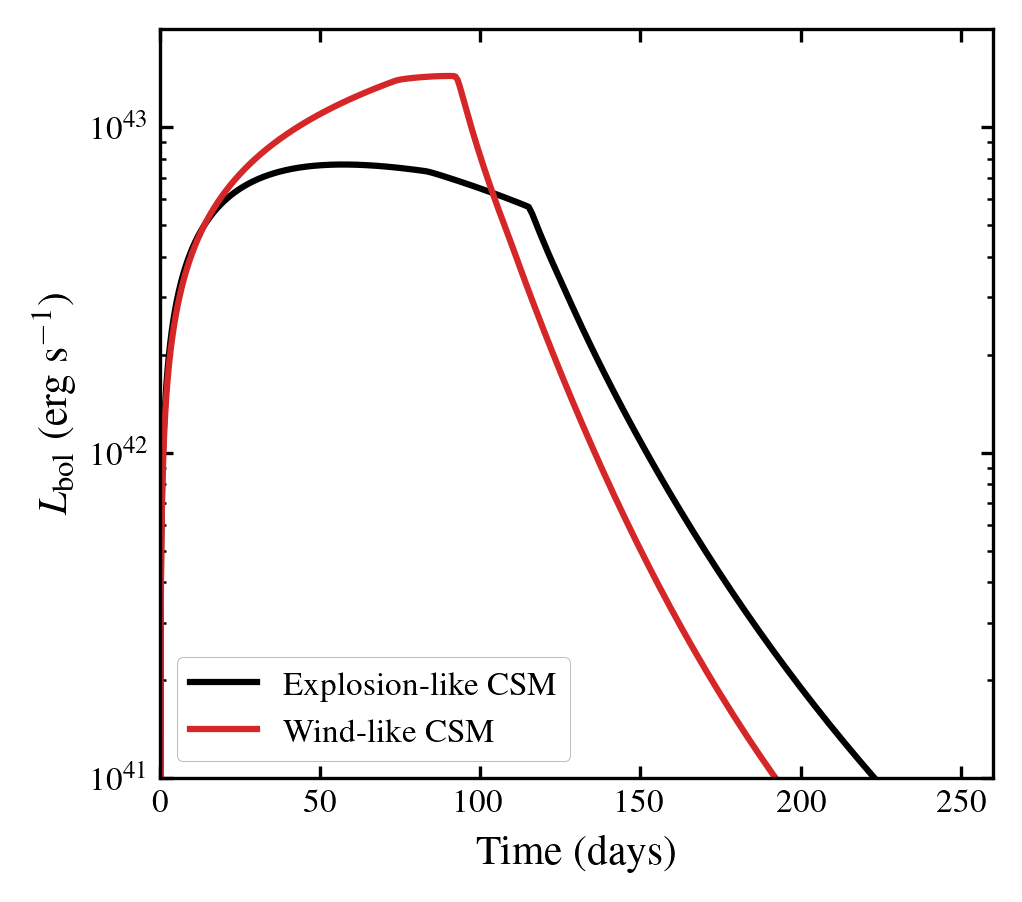}
\caption{Comparison of interaction-powered lightcurves for CSM models with identical density profiles at the time of explosion but different velocity structures. The black curve shows SN ejecta interacting with an explosion-like CSM undergoing homologous expansion ($v=r/t$). The SN ejecta has $M_{\rm ej}=5\,{\rm M_\odot}$ and $E_{\rm SN}=10^{51}$ erg, while the CSM eruption has $M_{\rm csm}=1\,{\rm M_\odot}$, $E_{\rm csm}=10^{49}$ erg, and an interval of $t_{\rm int}=1$ yr. The red curve corresponds to the same SN interacting with a wind-like CSM, constructed to have the same density distribution at the time of explosion but assigned a constant velocity of $300\,{\rm km\,s^{-1}}$. Both curves use the one-zone luminosity with $\kappa=0.2\,{\rm cm^2\,g^{-1}}$ and $\epsilon=0.5$.}
\label{fig:velocity_lightcurve}
\end{figure}

In Figure~\ref{fig:velocity_lightcurve}, we illustrate how differences in the assumed CSM velocity profile can significantly affect the resulting light curve. The black curve corresponds to a model in which the CSM was produced by an explosive event with a broken power-law density profile ($\delta = 0.5$, $n = 12$), with total mass $M_{\rm csm} = 1~{\rm M_\odot}$ and kinetic energy $E_{\rm csm}=10^{49}~{\rm erg}$. A SN explosion with the same broken power-law profile but with $M_\mathrm{ej} = 5~\mathrm{M}_\odot$ and $E_\mathrm{SN} = 10^{51}~\mathrm{erg}$ is launched into this CSM after an interaction delay of $t_\mathrm{int} = 1~\mathrm{yr}$.

For the red curve, we construct a CSM with exactly the same density profile at the time of SN explosion as in the black model, but assign it a constant velocity profile of $v = 300~\mathrm{km~s^{-1}}$. Despite the identical density structures at the onset of the SN, the resulting light curves differ markedly.

At early times ($\lesssim 20~\mathrm{d}$), the light curves are nearly identical, as the SN blast-wave velocity is much larger than the CSM velocity in both models. At intermediate times ($\gtrsim 20~\mathrm{d}$), differences in the velocity of the unshocked CSM at the shell location become increasingly important, causing the wind-like CSM model to be brighter than the explosion-like CSM model. At later times ($\gtrsim 90~\mathrm{d}$), the wind-like CSM model exhibits a rapid decline, while the explosion-like CSM model evolves more gradually. This behaviour arises because the outer regions of the homologously expanding CSM move to much larger radii at higher velocities than the assumed wind speed, delaying the time at which a given CSM mass shell interacts with the SN ejecta.

The quantitative differences demonstrated in this example show that the assumed CSM velocity profile can have a non-negligible impact on the resulting light curve, even when the density structure at the time of explosion is identical. Therefore, accurately accounting for the CSM velocity profile is essential when inferring CSM density distributions from observed interaction-powered SN light curves. This also raises the need for more careful scrutiny into the validity of inferences built on progenitor density profiles with models that can only capture constant velocity wind-like CSM.

\subsection{Aspherical CSM}
The spatial distribution of CSM can be highly aspherical, reflecting the diversity of mass-loss mechanisms. For instance, equatorially concentrated distributions or bipolar outflows are a natural consequence of binary interactions \citep[e.g.,][]{Morris2007, Kurfurst2020, Hirai2021,Orlando2024}. Observational evidence for such asphericity is abundant in interacting SNe, manifest in direct imaging \citep{Plait1995, Burrows1995}, complex multi-component spectral features \citep[e.g.,][]{Chugai1994, Fransson2002, Brethauer2022}, and polarimetry \citep[e.g.,][]{Leonard2000, Bilinski2018}.

To evaluate the impact of aspherical CSM on light curves, we implement a modular light curve combination framework. Our primary assumption is that the shell expansion is strictly radial, with negligible lateral deflection even in the presence of aspherical CSM. This simplification allows us to treat the shell propagation along each radial ray independently, analogous to a spherically symmetric case. Consequently, the integrated luminosity is calculated by summing the contributions from all radial rays.

\begin{figure}
\includegraphics[width=\columnwidth]{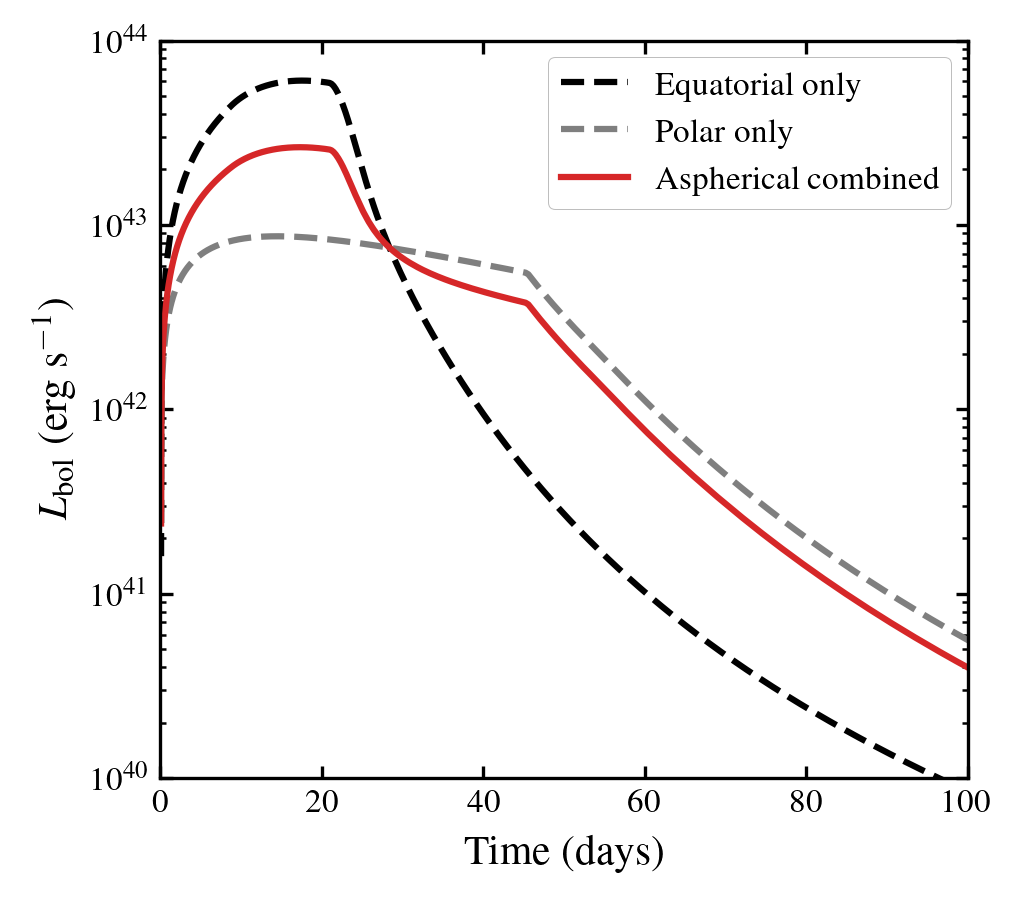}
\caption{Bolometric lightcurve for a SN with aspherical CSM. We model both the CSM and SN ejecta as explosions with broken power-law density distributions ($\delta=0.5$, $n=12$). The equatorial density distribution is equivalent to $M_{\rm expl}=1\,{\rm M_\odot}$ and $E_{\rm expl}=10^{49}$ erg, while the polar density distribution is equivalent to $M_{\rm expl}=0.2\,{\rm M_\odot}$ and $E_{\rm expl}=10^{49}$ erg. We assume a polar half-opening angle of $70$ deg and combine the independent radial lightcurves by solid angle. The SN has $M_{\rm expl}=2\,{\rm M_\odot}$ and $E_{\rm expl}=10^{51}$ erg; the luminosities use the same one-zone setup as the overview figure, with $\kappa=0.1\,{\rm cm^2\,g^{-1}}$ and $\epsilon=0.4$.}
\label{fig:2zone_lightcurve}
\end{figure}

As a representative case, we consider an axisymmetric, two-component CSM consisting of a dense equatorial torus and a more tenuous polar outflow, into which a spherically symmetric SN explosion is driven into. Fig.~\ref{fig:2zone_lightcurve} displays the resulting light curves: the dashed curves represent hypothetical spherically symmetric cases following the equatorial (black) and polar (grey) density profiles. Due to the density contrast, the shell propagates at disparate velocities, leading to distinct peak magnitudes and timescales.

The composite aspherical light curve (red solid curve) is a weighted average of these components. Notably, this light curve exhibits a double-peaked structure (at $\sim10$~d and $\sim30$~d). This feature could be misidentified as evidence for multiple discrete CSM shells if analysed under the assumption of spherical symmetry, thereby underscoring the necessity of accounting for aspherical CSM geometries in SN modelling.

We caution that this approach only works for optically thin CSM, as our diffusion treatment assumes spherical symmetry among other assumptions. For optically thick, aspherical CSM, the light curve requires proper multi-dimensional radiation hydrodynamics modelling \citep[e.g.][]{Chen2025}.

\subsection{Shell properties and Broadband photometry}
Although bolometric luminosities are useful, generally, we are limited to lightcurves seen in a handful of filters in wide-field surveys. Here we demonstrate additional properties from our numerical simulations which can provide further diagnostics that can be probed by these broadband lightcurves, and e.g., spectral observations. In particular, as we numerically solve Eq.~(\ref{eq:evolution1}) and Eq.~(\ref{eq:evolution2}), we store variables such as the shell mass, shell velocity, shell radius, and photospheric radius. We then approximate the spectral energy distribution (SED) as a blackbody with luminosity $L_{\rm obs}$, temperature
\begin{equation}
T_{\rm eff}(t)=\left(\frac{L_{\rm obs}(t)}{4\pi \sigma R_{\rm ph}(t)^2}\right)^{1/4},
\end{equation}
and photospheric radius $R_{\rm ph}$. This is a simplified assumption as real CSM interaction SEDs have rich spectral features. However, modelling the spectra requires complex radiative transfer simulations~\citep[e.g.,][]{Jerkstrand2015}, which is beyond the scope of this work. Assuming a blackbody is the standard approximation in other semi-analytical models for CSM interaction employed in light-curve fitting packages.

\begin{figure*}
\includegraphics[width=\textwidth]{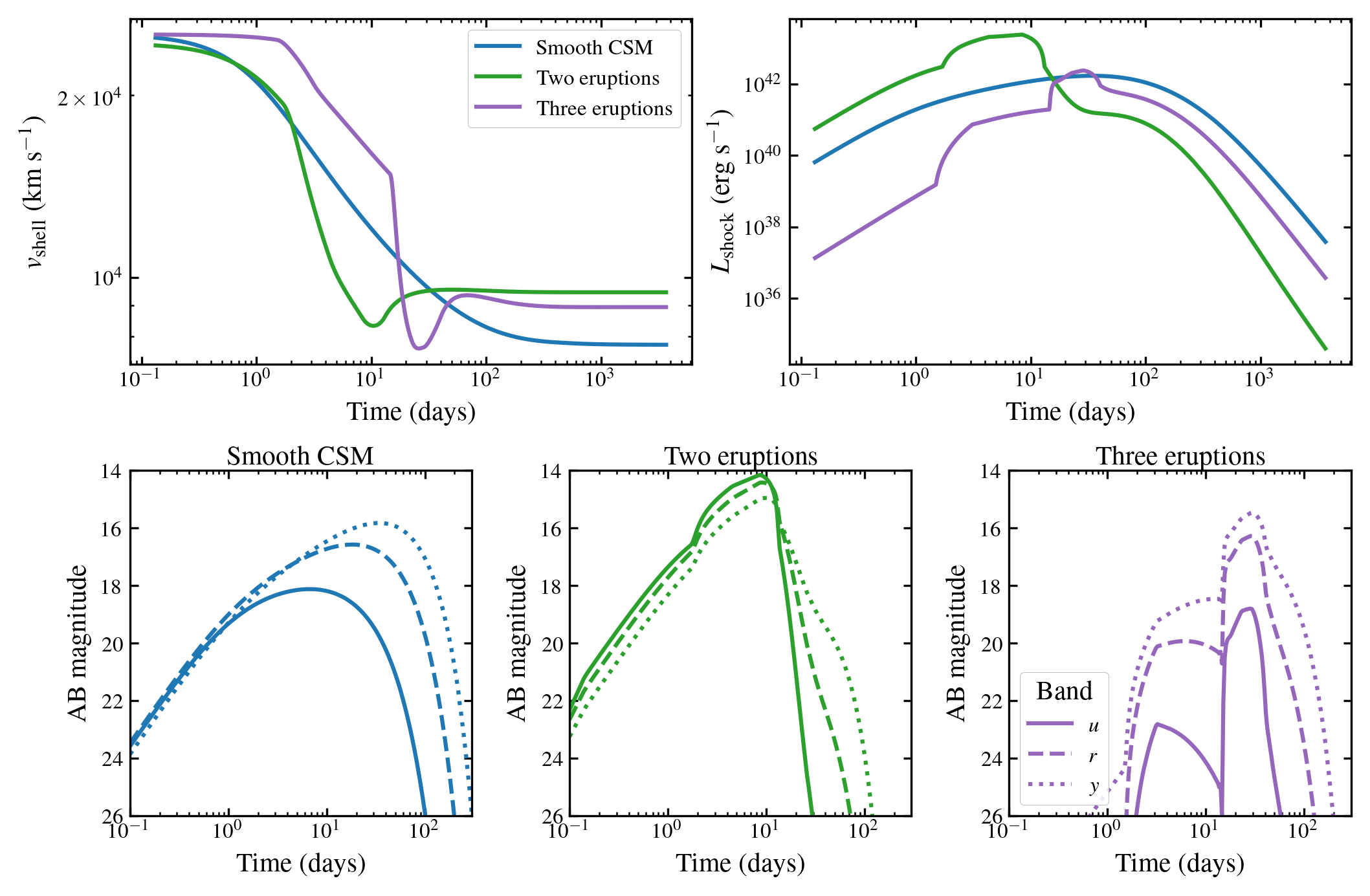}
\caption{Physical properties and LSST photometry for the representative multi-eruption CSM scenarios. \textbf{Top left:} Shell velocity evolution, showing the deceleration and re-acceleration caused by different CSM density structures. \textbf{Top right:} Raw shock luminosity before conversion to broadband magnitudes. \textbf{Bottom panels:} Synthetic LSST photometry in $u$ (solid), $r$ (dashed), and $y$ (dotted) bands at redshift $z=0.001$, assuming blackbody emission. Each lower panel corresponds to one of the CSM scenarios shown in Fig.~\ref{fig:multi_eruption_scenarios}: Smooth CSM (blue), Two eruptions (green), and Three eruptions (purple). All scenarios use a $2\,{\rm M_\odot}$, 1 foe broken power-law supernova, one-zone luminosities with $\kappa=0.1\,{\rm cm^2\,g^{-1}}$, and $\epsilon=0.4$.}
\label{fig:shell_properties_lsst}
\end{figure*}

In Figure~\ref{fig:shell_properties_lsst}, we show the shell velocity and raw shock luminosity in the top panels for the same three CSM scenarios shown in Fig.~\ref{fig:multi_eruption_scenarios}. The bottom panels show synthetic LSST photometry in the $u$, $r$, and $y$ bands, represented by solid, dashed, and dotted lines, respectively, at redshift $z = 0.001$. These magnitudes are calculated by approximating the SED as a blackbody with the luminosity and photospheric radius returned by the shock calculation.

Fig.~\ref{fig:shell_properties_lsst} reveals several features of the model. The shell velocities span the range expected for interaction-powered transients and respond directly to the CSM density structure: dense shells decelerate the interaction shell, while lower-density regions allow the shell to recover or evolve more smoothly. This time-dependent velocity is a useful diagnostic because spectroscopic velocities measured at different epochs need not trace the same physical layer of the interaction.

The raw shock luminosity panel shows where the dynamical interaction itself is strongest, before the luminosity is redistributed into broadband colours by the blackbody approximation. The lower panels then show how the same CSM structures would appear in LSST-like photometry. Some density features map cleanly into photometric bumps, while others are muted or appear mainly in one band because the observed magnitudes depend on both luminosity and temperature. This is useful but also a caution: a smooth optical lightcurve does not necessarily imply a smooth CSM, especially when temperature evolution and cadence effects wash out structure.

For these fiducial supernova parameters, the models are bright enough to be detectable by LSST over a substantial fraction of their evolution. The detailed colour evolution carries information about the temperature history, while the timing of bumps or shoulders traces the radii of dense CSM structures. However, resolving these features will require cadence faster than the characteristic shock-crossing time of the shells, and in some cases complementary faster surveys such as ZTF~\citep{Bellm2019} and LS4~\citep{Miller2025} will be more important than depth alone.

\section{Multi-wavelength signatures}\label{sec:multi}
The same shock trajectory used for the optical lightcurves also provides radio and X-ray diagnostics. The aim here is not to replace detailed non-thermal or plasma calculations, but to keep the optical, radio, and X-ray estimates tied to the same CSM density, shock radius, and shock velocity.

\subsection{Synchrotron emission}

\begin{figure*}
\includegraphics[width=\textwidth]{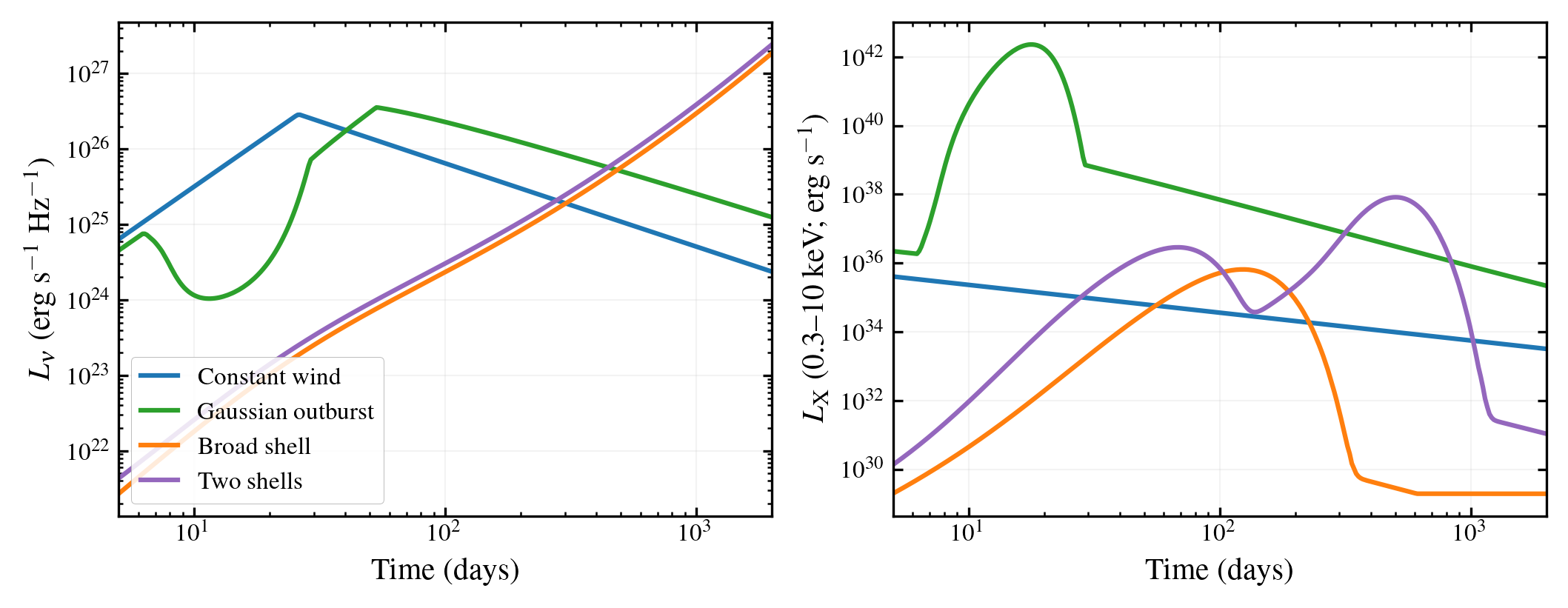}
\caption{Self-consistent radio and X-ray diagnostics for four representative CSM scenarios. The left panel shows 5 GHz synchrotron spectral luminosities, $L_\nu$, while the right panel shows intrinsic 0.3--10 keV thermal bremsstrahlung (free-free) luminosities. The four curves correspond to a constant wind (blue; $\dot{M}=10^{-4}\,{\rm M_\odot\,yr^{-1}}$, $v_w=500\,{\rm km\,s^{-1}}$), a Gaussian outburst in the mass-loss history (green; baseline $\dot{M}=10^{-5}\,{\rm M_\odot\,yr^{-1}}$, peak $\dot{M}=0.1\,{\rm M_\odot\,yr^{-1}}$, $v_w=100\,{\rm km\,s^{-1}}$), a broad CSM shell (orange; a power-law background plus one broad shell at $1.2\times10^{16}$ cm), and a two-shell CSM snapshot (purple; shells at $7\times10^{15}$ cm and $4\times10^{16}$ cm). These generic shell examples are not the same CSM profiles as the multi-eruption optical examples in Fig.~\ref{fig:multi_eruption_scenarios}; they are chosen to illustrate how radio and X-ray diagnostics respond to different density structures. All curves use the same $2\,{\rm M_\odot}$, 1 foe broken power-law supernova and $\epsilon=0.4$. The radio calculation adopts $\epsilon_B=0.01$, $\epsilon_e=0.1$, and $p=3$. The X-ray panel uses the same shock solution and a fast emission-measure thermal free-free estimate, capped by the available shock power and evaluated with a fiducial emitting-volume/filling-factor scale of 0.1; no absorption is applied in this illustrative figure.}
\label{fig:multiwavelength}
\end{figure*}

In addition to optical and UV emission from the shock-heated CSM, SN-CSM interactions produce characteristic signatures at radio wavelengths. Radio observations provide complementary diagnostics of the shock velocity, magnetic field amplification, and CSM density structure that are largely independent of the optical emission.

Radio emission arises from synchrotron radiation produced by relativistic electrons accelerated in the forward shock, gyrating in amplified magnetic fields. Following the formalism of \citet{Chevalier1998} and as previously implemented in \program{Redback}~\citep{Sarin2024}, we calculate the synchrotron flux density as
\be
L_\nu = \frac{C_0}{2 \nu_L} \left(\frac{\nu}{\nu_L}\right)^{(1-p)/2},
\ee
where $C_0 = \frac{4}{3} N_0 \sigma_T c u_B$ is the synchrotron power, $N_0 = \frac{4\pi}{3} r_s^3 \epsilon_e n_{\rm csm}$ is the number of radiating electrons, $\nu_L = eB/(2\pi m_e c)$ is the Larmor frequency, and $p$ is the electron power-law index. The key microphysical parameters are $\epsilon_B$ (magnetic field energy fraction, where $B = \sqrt{8\pi \epsilon_B \rho_{\rm csm} v_s^2}$), $\epsilon_e$ (fraction of shock energy in relativistic electrons), and $p$ (typically $\sim$3). At low frequencies, synchrotron self-absorption becomes important, with the SSA frequency given by
\be
\nu_{\rm SSA} = \left(\frac{d_L^2 \cdot 3^{1.5} \cdot e^{0.5} B^{0.5} F_{\nu,0} \nu_L^{\beta - 1}}{4\pi^{1.5} r_s^2 c^{0.5} m_e^{1.5}}\right)^{2/(2\beta + 3)},
\ee
where $\beta = 1 - (1-p)/2$ and $F_{\nu,0} = C_0/(2\nu_L)/(4\pi d_L^2)$. For $\nu < \nu_{\rm SSA}$, the spectrum transitions to $F_\nu \propto \nu^{2.5}$. 
These expressions help provide some scalings, in particular, collecting the $\epsilon_B$ dependence, $C_0 \propto u_B \propto \epsilon_B$ and $\nu_L \propto B \propto \epsilon_B^{1/2}$, gives an overall scaling $L_\nu \propto \epsilon_B^{(p+1)/4}$, which is positive for all $p > -1$, i.e., stronger magnetic fields always increase the radio luminosity. Meanwhile, the full scaling with shock properties is $L_{\rm radio} \propto n_{\rm csm} v_s^2 r_s^3 \epsilon_B^{(p+1)/4} \epsilon_e$. For a wind-like CSM profile where $\rho \propto r^{-2}$, this gives $L_{\rm radio} \propto r_s v_s^2$. The same synchrotron formalism as implemented in \program{Redback} can also be evaluated at X-ray frequencies. Although for much of the parameter space this emission component is dimmer in X-rays than other emission mechanisms such as inverse-Compton or bremsstrahlung as we discuss further below.

The left panel of Fig.~\ref{fig:multiwavelength} shows the radio spectral luminosity at 5 GHz for four representative CSM scenarios calculated from the same shock model. We adopt synchrotron parameters $\epsilon_B=0.01$, $\epsilon_e=0.1$, and $p=3$, typical for SN shocks~\citep{Chevalier2017}. A steady wind produces a relatively smooth radio evolution because the shock sees a monotonic density profile. By contrast, a Gaussian outburst or discrete shells can delay or enhance the radio luminosity when the shock reaches the denser material. The radio curves therefore provide a complementary view of the same CSM structures shown in the optical examples, but with different sensitivity to shock radius, shock velocity, magnetic-field amplification, and absorption.
We again stress that this is not a detailed radio spectral model, it neglects reverse-shock emission, external free-free absorption, inverse-Compton cooling, and detailed electron-energy evolution~\citep{Margalit2021}. Instead, it provides a fast, self-consistent check of whether an optical CSM solution should also be radio bright, and whether radio observations are compatible with the inferred density and velocity structure.

\subsection{Thermal bremsstrahlung X-ray emission}
In dense SN-CSM interactions, X-ray emission is often expected from thermal bremsstrahlung in the hot, shocked gas, with post-shock temperatures $T \sim 10^8$ K for typical shock velocities $v_s \sim 10^4$ km s$^{-1}$~\citep{Chevalier2017, Margalit2022_xrays}. At very high shock velocities or lower densities, non-thermal synchrotron or inverse-Compton emission can instead become important. In principle, this emission provides a sensitive probe of the CSM density structure due to its quadratic density dependence ($L_{\rm X} \propto n_{\rm csm}^2$), making it particularly diagnostic of high-density shells and clumps~\citep{Chandra2018}. 

Here we implement a fast post-processing estimate of the thermal free-free luminosity using the shock trajectory and upstream CSM density already calculated by the optical model. We estimate the post-shock temperature from the strong-shock relation
\begin{equation}
kT_{\rm sh} = \frac{3}{16}\mu m_{\rm p} v_{\rm sh}^2,
\end{equation}
allowing for an optional electron-temperature factor to account for incomplete electron-ion equilibration. The intrinsic free-free luminosity is then calculated from
\begin{equation}
L_{\rm ff} = 1.426\times10^{-27} g_{\rm ff} T_{\rm e}^{1/2} \int n_{\rm e}n_{\rm i}\,dV,
\end{equation}
where $g_{\rm ff}$ is the Gaunt factor and $T_{\rm e}=f_{\rm e,T}T_{\rm sh}$ is the electron temperature, with an optional electron-ion equilibration factor, $f_{\rm e,T}$, which we set to unity. Because the thin-shell model does not resolve the detailed X-ray cooling layer, we estimate the integral from the swept-up CSM mass and the post-shock density $\rho_{\rm post}=\chi\rho_{\rm csm}$, where $\chi=4$ by default is the strong-shock jump-condition compression factor for an ideal gas with $\gamma=5/3$ and can be varied in the code,
\begin{equation}
\int n_{\rm e}n_{\rm i}\,dV \simeq
\frac{M_{\rm csm,sw}\rho_{\rm post}}{\mu_{\rm e}\mu_{\rm i}m_{\rm p}^2}.
\end{equation}
The band-limited luminosity is obtained by integrating the thermal bremsstrahlung spectrum over a requested X-ray band (which we set by default to 0.3--10 keV).

As the integral estimate above is a thin-shell post-processing approximation, we also impose a local energy-budget cap on the bolometric free-free luminosity,
\begin{equation}
L_{\rm X,bol} =
\min\left(L_{\rm ff,bol},\, f_{\rm X,max} L_{\rm shock}\right),
\end{equation}
where $L_{\rm shock}$ is the selected forward, reverse, or total shock-power scale and $f_{\rm X,max}$ is an input cap. This prevents the unresolved cooling-layer estimate from radiating more X-ray power than is available from the shock.

Soft X-rays from dense CSM interaction can be strongly absorbed. We therefore include multiplicative photoelectric absorption from Milky Way, host-galaxy, and optional local CSM columns. The local CSM column is only an approximate upstream column estimate; for detailed comparisons to X-ray spectra it should be replaced by a column calculated from the inferred CSM structure and ionisation state. We also allow the user to select the forward shock, reverse shock, or total shock component as the power scale, although the emission-measure estimate is most self-consistent when coupled to the forward-shock luminosity, because it uses the upstream CSM density and swept-up CSM mass.

The right panel of Fig.~\ref{fig:multiwavelength} shows the corresponding 0.3--10 keV thermal bremsstrahlung luminosities for the same shock trajectories used in the radio panel. The differences between the radio and X-ray morphology are therefore driven by the different density, velocity, and optical-depth scalings of the two emission processes, rather than by changing the underlying dynamical model. We stress that this X-ray calculation should be regarded with more caution than the radio calculation. The observed X-ray luminosity depends sensitively on the ionisation state, non-equilibrium electron heating, clumping, line cooling, and absorption in the unshocked CSM. Inverse Compton emission and non-thermal synchrotron emission can also contribute in some regimes~\citep{Chandra2018}. The value of the current implementation is therefore not that it provides a definitive X-ray prediction, but that it gives a fast, physically motivated estimate of whether the same CSM structure inferred from the optical lightcurve should plausibly be bright or hidden in X-rays, especially in the regime where one would expect thermal bremsstrahlung emission.
%%%%%%%%%%%%%%%%%%%%%%%%%%%%
\section{Inference}\label{sec:inference}
%%%%%%%%%%%%%%%%%
\subsection{Recovering variable wind mass-loss histories}\label{sec:variable_wind_inference}

To validate our model and framework's ability to constrain CSM properties, we perform an inference study using synthetic data generated from a Gaussian wind model. This scenario is physically motivated by assuming that the star undergoes a period of enhanced wind mass-loss a few decades before core-collapse, as seen in numerous simulations of massive stars~\citep[e.g.][]{Quataert2012,Smith2014}.  Both observations of eruptive mass loss and theoretical models of wave-driven or explosive pre-SN ejection suggest that the outflow velocity need not remain fixed during enhanced mass-loss episodes~\citep[e.g.][]{Vink1999,Groh2011}. In this recovery experiment we deliberately keep $v_{\rm wind}$ constant so that the inferred density scale can be compared cleanly to the injected $\dot{M}/ v_{\rm wind}$.
The synthetic data are generated and recovered with the same one-zone luminosity calculation, so that this exercise validates the inference machinery and the recovery of an effective mass-loss history rather than mixing different luminosity prescriptions.

We first generate synthetic bolometric observations by modelling a Type IIn-like supernova interacting with a circumstellar wind characterized by a Gaussian enhancement in mass-loss rate. The mass-loss history is parametrized as:
\begin{equation}
  \dot{M}(t) = \dot{M}_{\rm baseline} + (\dot{M}_{\rm peak} - \dot{M}_{\rm baseline})
  \exp\left[-\frac{(t - t_{\rm peak})^2}{2\sigma_t^2}\right],
\end{equation}
where $t$ represents the time before explosion, $t_{\rm peak}$ is the epoch of peak mass loss, $\sigma_t$ is the temporal width of the enhancement, $\dot{M}_{\rm baseline}$ is the quiescent wind mass-loss rate, and $\dot{M}_{\rm peak}$ is the peak mass-loss rate
during the outburst. Following our CSM interface implementation
(Section~\ref{sec:model}), we discretize this continuous profile into 50 temporal bins spanning $t_{\rm peak} \pm 4\sigma_t$, which is then used to construct a density profile assuming $\rho \propto \dot{M}r^{-2}$.

For our fiducial test case, we adopt $t_{\rm peak} = 60$~years, $\sigma_t = 25$~years, $\dot{M}_{\rm baseline} = 10^{-5}$~M$_\odot$~yr$^{-1}$, $\dot{M}_{\rm peak} = 8 \times 10^{-5}$~M$_\odot$~yr$^{-1}$, and a constant wind velocity $v_{\rm wind} = 80$~km~s$^{-1}$. The supernova ejecta are modeled with an
exponential density profile with $M_{\rm ej} = 3.0$~M$_\odot$ and $E_{\rm SN} = 1.5$~foe. We assume a radiative efficiency $\epsilon = 0.3$ and an opacity of $\kappa = 0.1$~cm$^2$~g$^{-1}$. We sample this model at 35 logarithmically-spaced epochs between 1 and 500~days and add Gaussian noise with a fractional uncertainty of 12\%, representative of the reconstructed bolometric luminosity of a well-sampled supernova.

We perform Bayesian inference using the \textsc{pymultinest}~\citep{Feroz2009, Buchner2016} nested sampler as implemented in \textsc{Redback}~\citep{Sarin2024} to recover the input parameters from the synthetic lightcurve. We use log-uniform priors on $t_{\rm peak} \in [10, 200]$~years and $\sigma_t \in [5, 100]$~years, log-uniform priors on $\dot{M}_{\rm baseline} \in [10^{-6}, 10^{-4}]$~M$_\odot$~yr$^{-1}$ and $\dot{M}_{\rm peak} \in [10^{-6}, 10]$~M$_\odot$~yr$^{-1}$, and log-uniform priors on $v_{\rm wind} \in [10, 300]$~km~s$^{-1}$, $M_{\rm ej} \in [1, 30]$~M$_\odot$, and $E_{\rm SN} \in [0.1, 10]$~foe, with a uniform prior on $\epsilon \in [0.01, 1]$. The opacity is fixed to its true value $\kappa = 0.1$~cm$^2$~g$^{-1}$.

\begin{figure}
\includegraphics[width=\columnwidth]{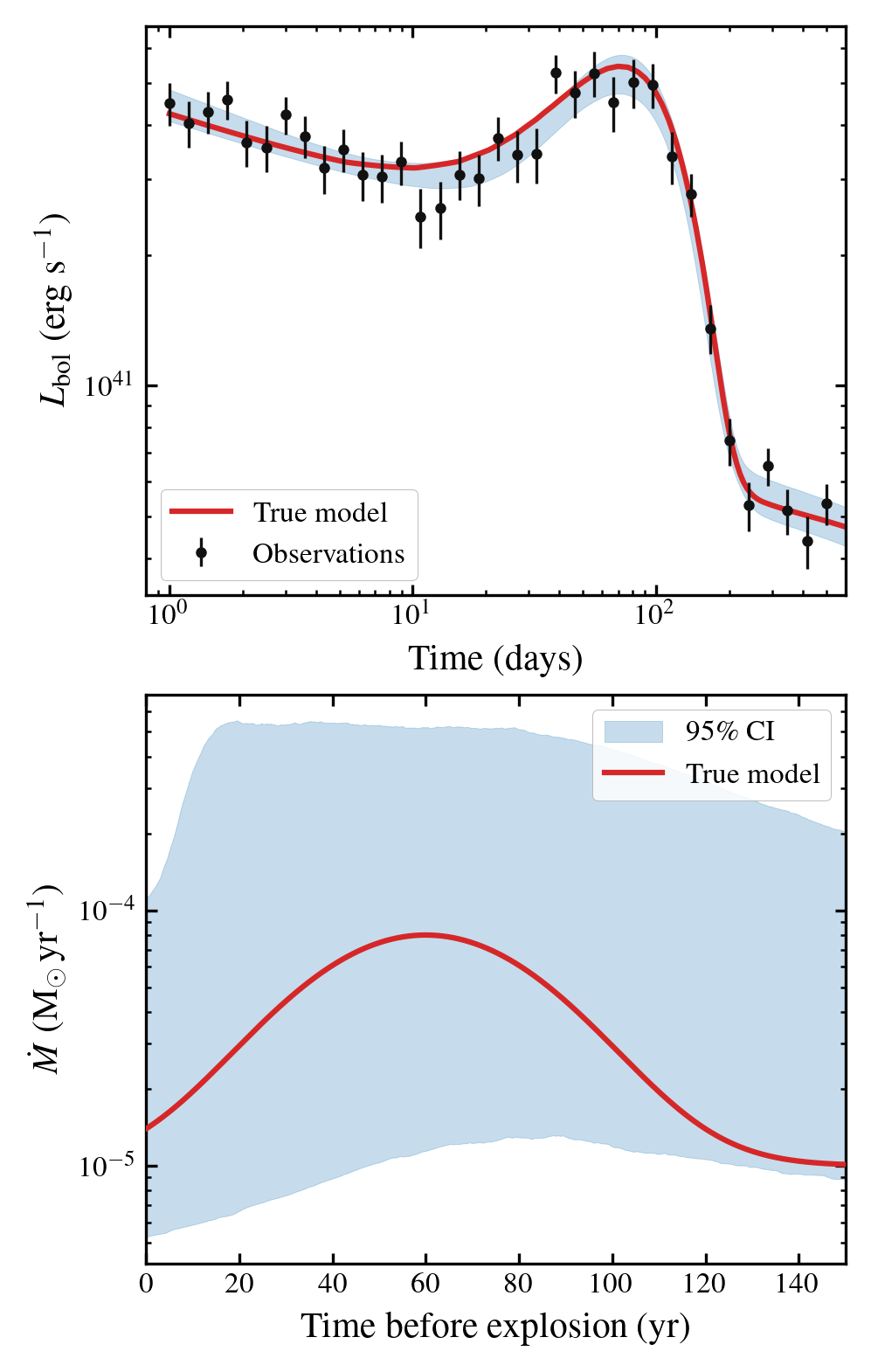}
\caption{\textbf{Top:} Posterior predictive lightcurve for the Gaussian wind inference test. Data points show synthetic observations with 12\% fractional uncertainty. The red line shows the true model and the blue shaded region the 95\% posterior credible interval. \textbf{Bottom:} Recovered mass-loss history. The red line shows the true Gaussian wind profile and blue shaded regions the 95\% posterior credible interval.
Despite significant degeneracies among individual parameters (Table~\ref{tab:recovery}), the combined posterior encompasses the true mass-loss history, demonstrating that photometric data constrains the effective CSM density structure.}
\label{fig:gaussian_wind_inference}
\end{figure}

Figure~\ref{fig:gaussian_wind_inference} presents our inference results and Table~\ref{tab:recovery} summarises the parameter recovery. The timing parameters are well recovered: $t_{\rm peak} = 62.6^{+61.6}_{-35.5}$~years and $\sigma_t = 23.9^{+23.1}_{-13.4}$~years, both consistent with the injected values. Credible intervals here and throughout are quoted at the $68\%$ level unless otherwise stated.

The mass-loss rates and wind velocity exhibit significant degeneracies. While $\dot{M}_{\rm peak}$ and $v_{\rm wind}$ individually show broad posteriors offset from the true values, the more directly photometrically constrained combination $\dot{M}_{\rm peak}/v_{\rm wind}$, which directly sets the CSM density $\rho_{\rm csm} =
\dot{M}/(4\pi r^2 v_{\rm wind})$, is well constrained and encompasses the true value within the 95\% credible interval. Bolometric photometry alone mainly constrains their ratio for a wind-like CSM. If $v_{\rm wind}$ is measured independently from narrow spectral features, pre-explosion eruption timing, or other multi-wavelength diagnostics, the inferred mass loading can be converted into a a more robust mass-loss rate measurement. Similarly, while $E_{\rm SN}$ and $\epsilon$ are individually offset, their product $E_{\rm SN} \times \epsilon$, which governs the luminosity normalization, is recovered within the 95\% credible interval. These are direct degeneracies of CSM interaction physics: photometric data alone constrains the effective CSM density structure and energy input, but cannot uniquely separate the
individual parameters without additional constraints such as spectroscopic shock velocities or indirect mass-loss constraints from pre-explosion imaging.

Despite these parameter-level degeneracies, the posterior credible intervals in Figure~\ref{fig:gaussian_wind_inference} encompass the true mass-loss profile across its full extent, demonstrating that the inferred CSM density structure and observable lightcurve are well determined even when individual parameters are not. This also stresses the need to consider the full posterior as opposed to individual point estimates which will likely miss the true input. 

\begin{table}
  \centering
  \caption{Parameter recovery for the Gaussian wind inference test. Quoted values are the posterior median with $68\%$ credible intervals. Parameters marked with $\dagger$ are individually degenerate; their physically constrained combinations ($\dot{M}_{\rm peak}/v_{\rm wind}$ and $E_{\rm SN}\times\epsilon$) are each recovered within the $95\%$ credible interval.}
  \label{tab:recovery}
\begin{tabular}{llcc}
      \hline
      Parameter & Symbol & True & Recovered ($68\%$ CI) \\
      \hline
      Peak time (yr)
        & $t_{\rm peak}$
        & $60$
        & $62.6^{+61.6}_{-35.5}$ \\
      Width (yr)
        & $\sigma_t$
        & $25$
        & $23.9^{+23.1}_{-13.4}$ \\
      Baseline $\dot{M}$ (M$_\odot$ yr$^{-1}$)
        & $\dot{M}_{\rm baseline}$
        & $1.0\times10^{-5}$
        & $2.9^{+3.8}_{-2.0}$$\times10^{-5}$ \\
      Peak $\dot{M}^{\dagger}$ (M$_\odot$ yr$^{-1}$)
        & $\dot{M}_{\rm peak}$
        & $8.0\times10^{-5}$
        & $2.1^{+2.7}_{-1.4}$$\times10^{-4}$ \\
      Wind velocity$^{\dagger}$ (km s$^{-1}$)
        & $v_{\rm wind}$
        & $80$
        & $36.3^{+66.1}_{-20.3}$ \\
      Ejecta mass (M$_\odot$)
        & $M_{\rm ej}$
        & $3.0$
        & $2.6^{+3.4}_{-1.2}$ \\
      Explosion energy$^{\dagger}$ (foe)
        & $E_{\rm SN}$
        & $1.5$
        & $0.38^{+0.35}_{-0.16}$ \\
      Efficiency$^{\dagger}$
        & $\epsilon$
        & $0.3$
        & $0.65^{+0.24}_{-0.29}$ \\
      \hline
  \end{tabular}
\end{table}

%%%%%%%%%%%%%%%%%
\subsection{Analysis of real transients}

\begin{figure*}
\includegraphics[width=0.95\textwidth]{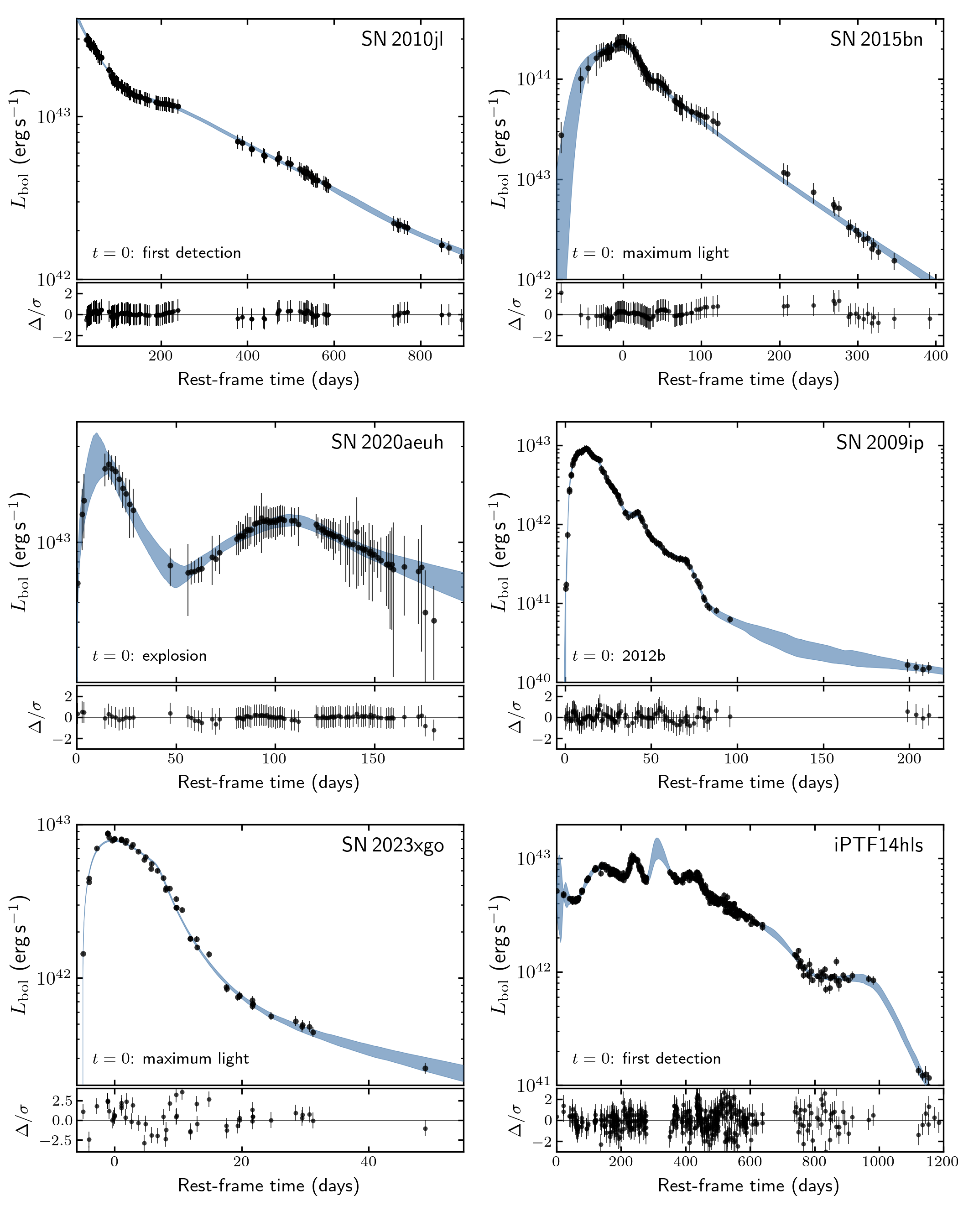}
\caption{Bolometric luminosity data and our fits to the lightcurve for all six real transients analysed in this work. The blue bands show the 90\% credible interval. The lower panel in each object shows $(L_{\rm data}-L_{\rm model})/\sigma$ relative to the maximum-likelihood curve, using the same uncertainties as the likelihood. The x-axis is shown as rest-frame time in days; the zero point follows the convention of the underlying bolometric data and is noted in each panel: first detection for SN~2010jl and iPTF14hls, maximum light for SN~2015bn and SN~2023xgo, explosion for SN~2020aeuh, and the 2012b event for SN~2009ip.}
\label{fig:real_lcs}
\end{figure*}

Having validated our framework on synthetic observations where we knew the true input, we now apply our models to interpret a range of real and likely interaction-powered transients to test our model in different regimes. We discuss the objects in the same order as Fig.~\ref{fig:real_lcs}: SN~2010jl, a luminous Type IIn with sustained interaction over $\sim 900$ days requiring significant wind mass loss; SN~2020aeuh, a Type Ia-CSM where we disentangle thermonuclear and interaction contributions; SN~2023xgo, a rapidly-evolving Type Ibn/Icn that probes the stripped-envelope progenitor regime with fast winds; SN~2015bn, a superluminous supernova where we test whether continuous wind or discrete shell parameterisations converge on the same physical picture; SN~2009ip, an LBV progenitor with extensively documented pre-SN eruptions allowing direct comparison between CSM model and known activity history; and iPTF14hls, an extraordinary long-duration event where we compare discrete-shell and p-spline CSM reconstructions for its $\sim 600$-day plateau and repeated brightness fluctuations.

Together, these six objects span a large diversity of interaction-powered transients: from hydrogen-rich core-collapse (SN~2010jl, and potentially iPTF14hls) to hydrogen-poor stripped-envelope (SN~2023xgo) to thermonuclear
(SN~2020aeuh) explosions; from smooth wind-dominated CSM (SN~2010jl, SN~2023xgo) to flexible density reconstructions and structured CSM (SN~2009ip, iPTF14hls); from single-eruption scenarios (SN~2015bn) to multi-eruption histories (SN~2009ip, iPTF14hls); and from relatively ordinary luminosities ($\sim 10^{42}$~erg~s$^{-1}$) to superluminous events ($\sim 10^{44}$~erg~s$^{-1}$). This sample thus provides a comprehensive test of our framework's flexibility and ability to recover physically meaningful parameters across different progenitor channels, mass-loss mechanisms, and explosion types.

In Fig.~\ref{fig:real_lcs}, we show the bolometric lightcurve data alongside the inferred model fits, and in Fig.~\ref{fig:real_densities} the corresponding mass-loss histories or density profiles. Where more than one CSM parameterisation was fitted to an object, the figures show the model that is either statistically preferred or most physically interpretable; showing every alternative fit would make the already multi-object figure difficult to read. We discuss the relevant alternative fits in the object-by-object text below. For the real-transient inference in this section we use the fast one-zone lightcurve calculation, with opacity supplied where appropriate so that the shock luminosity is diffusion processed. This choice is deliberate: several of the fitted CSM structures are homologously expanding or phenomenological density snapshots rather than finite, static shells, and therefore fall outside the strict domain of the transport-mode calculation discussed in Sec.~\ref{sec:model}.
The parameters inferred from bolometric lightcurves should also be read with the main thin-shell degeneracies in mind. The dynamics and luminosity depend most directly on an effective CSM density scale, such as $\dot{M}/v_{\rm wind}$ for wind models, and on the product of shock efficiency and available kinetic power. For this reason, Fig.~\ref{fig:real_densities} and the discussion below should be interpreted primarily as constraints on the density or density-equivalent mass-loading history when only bolometric photometry is used. We retain $\dot{M}(t)$ in the figures because it is the parameter used by the wind models and is easier to compare with stellar-evolution predictions, but these histories are conditional on the adopted velocity prior unless $v_{\rm wind}$ is independently constrained. Narrow-line velocities, pre-explosion eruption dates combined with shell radii, radio/X-ray absorption columns, or spectroscopic shock velocities can all help break this degeneracy. Therefore, wind velocities, mass-loss rates, CSM masses, and efficiencies should be read as posterior summaries within the adopted parameterisation and priors, with their physical interpretation strengthened when external velocity, geometry, or multi-wavelength constraints are available. This is particularly important for generic shell and p-spline profiles, where the quoted CSM masses are spherical-equivalent density integrals. If the material covers only a fraction $f_\Omega$ of the sky as seen from the progenitor, and has an effective volume filling factor $f_{\rm fill}$, then the physical mass should be read approximately as
\begin{equation}
M_{\rm csm,true} \simeq f_\Omega f_{\rm fill} M_{\rm csm,sph}.
\end{equation}
Without independent constraints on geometry or clumping, we therefore interpret these masses as density-integral normalisations rather than literal isotropic ejecta masses.
Throughout this section, unless specified otherwise, we fit the data using \program{Redback}~\citep{Sarin2024}, sampling the posterior through the \program{pymultinest}~\citep{Feroz2009, Buchner2016} sampler, implemented in \program{Bilby}~\citep{Ashton2019}. 
For the high-dimensional spline reconstructions of SN~2009ip and iPTF14hls we instead use local \program{emcee} posterior sampling initialised around maximum-likelihood p-spline density reconstructions. We also exclusively fit the bolometric luminosities to avoid any systematics associated with an incorrect characterisation of the spectral energy distribution. However, we note that while our chosen supernovae are extensively observed, their bolometric lightcurves may still not be truly bolometric.
For completeness, Appendix~\ref{app:realfit_tables} gives the priors and posterior summaries for the adopted real-transient fits shown in Figs.~\ref{fig:real_lcs} and~\ref{fig:real_densities}. The full prior files and plotting scripts are also released with the software.
%%%%%%%%%%%%%%%%%%%%%%%%%%%%
\subsubsection{SN~2010jl}
SN~2010jl is one of the most luminous and best-studied Type IIn supernovae, discovered in UGC~5189A in November 2010~\citep{Newton2010}. The event exhibited sustained high luminosity for over 900 days, with a peak bolometric luminosity of $\sim 3 \times 10^{43}$~erg~s$^{-1}$ that remained above $10^{42}$~erg~s$^{-1}$ even at 850 days post-discovery~\citep{Stoll2011, Smith2011, Zhang2012}. The total radiated energy exceeded $6.5 \times 10^{50}$~erg, placing it among the most energetic core-collapse supernovae observed~\citep{Fransson2014}.

Previous studies have established that SN~2010jl's extraordinary luminosity is powered by interaction with dense CSM. \citet{Fransson2014} analysed HST and ground-based observations spanning up to 1100 days post first light, finding the light curve consistent with a radiative shock propagating through an $r^{-2}$ CSM with mass-loss rate $\dot{M} \sim 0.1$~M$_\odot$~yr$^{-1}$ and total CSM mass $\gtrsim 3$~M$_\odot$. 
Spectroscopic observations revealed narrow emission components with expansion velocities of $\sim 100$~km~s$^{-1}$, characteristic of LBV winds. \citet{Ofek2014} reached similar conclusions through X-ray and optical analysis, while \citet{Chandra2015} detected radio emission consistent with dense CSM at larger radii. These are also consistent with independent analyses of the spectra and lightcurve with radiation hydrodynamical modelling~\citep{Dessart2015}.
The collective evidence strongly favours an LBV-like progenitor that underwent decades of enhanced mass loss before explosion. 

We fitted the bolometric light curve data from~\citet{Fransson2014} (observations spanning 28 to 901 days post first detection), under the scenario of a variable wind mass-loss to allow the model to explore the possibility of a variable instead of constant mass-loss history. We parameterise the mass-loss history as a smoothly connected triple power-law with broad priors on break times (years before supernova),
\begin{equation}
\dot{M}(t) = \dot{M}_0 \times
\begin{cases}
t^{\alpha_1} & t < t_{\rm break,1} \\
t_{\rm break,1}^{\alpha_1}\left(\dfrac{t}{t_{\rm break,1}}\right)^{\alpha_2} & t_{\rm break,1} \leq t < t_{\rm break,2} \\
t_{\rm break,1}^{\alpha_1}\left(\dfrac{t_{\rm break,2}}{t_{\rm break,1}}\right)^{\alpha_2}\left(\dfrac{t}{t_{\rm break,2}}\right)^{\alpha_3} & t \geq t_{\rm break,2},
\end{cases}
\label{eq:mdot_powerlaw}
\end{equation}
where $t$ is time before explosion in years. We adopt broad overlapping priors on mass-loss rates and power-law indices such that if the data is well described by a constant or minimally changing mass-loss, this could be a viable solution for the sampler. The full prior ranges and posterior credible intervals for this fit are summarised in Appendix~\ref{app:realfit_tables}. We fix $\kappa = 0.34$~cm$^2$~g$^{-1}$ given the Type II classification.

\begin{figure*}
\includegraphics[width=\textwidth]{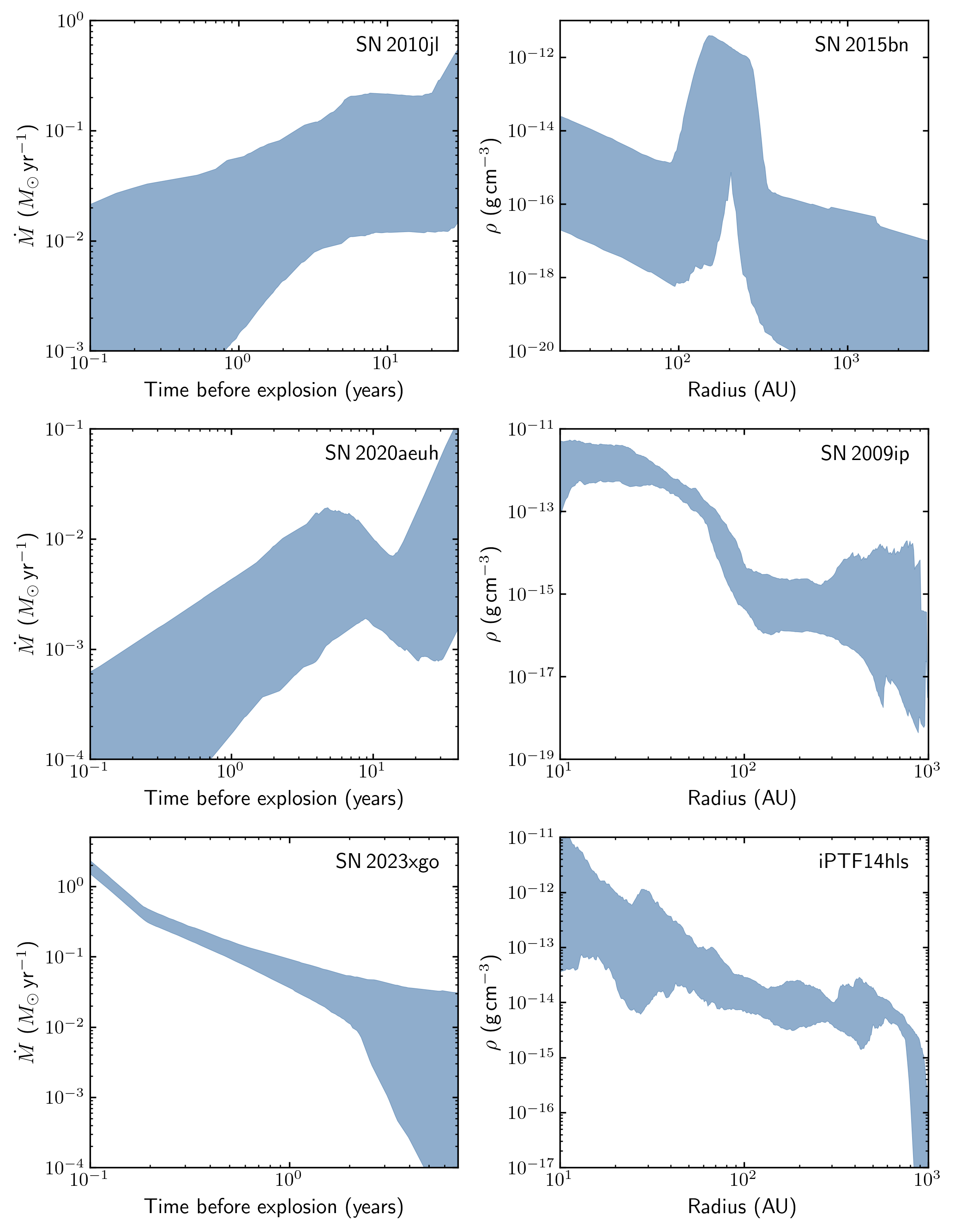}
\caption{Inferred mass-loss histories or density profiles at the time of the supernova (90\% credible interval based on the lightcurve fits) for all six real transients analysed in this work.}
\label{fig:real_densities}
\end{figure*}

Our fit to the light curve shows good agreement with the data, indicating that our framework is able to capture the interaction seen in SN~2010jl. As shown in Fig.~\ref{fig:real_densities}, we find support for a time-varying mass-loss history with distinct phases. Our inference yields $\dot{M}_0 = 7.8^{+24.5}_{-5.3}\times10^{-3}\,M_\odot\,\mathrm{yr}^{-1}$ at the reference time of $1$\,yr before supernova. The mass-loss history exhibits distinct phases characterised by different power-law indices: a decline toward explosion in the final years in our lookback-time convention ($\alpha_1 = 0.97^{+0.59}_{-0.44}$), a nearly flat phase at intermediate times ($\alpha_2 = -0.05^{+0.09}_{-0.08}$), and another time-decaying mass-loss phase in the distant past, corresponding to higher mass loss at larger lookback times ($\alpha_3 = 2.10^{+0.25}_{-0.19}$), with transitions at $t_{\rm break,1} = 6.5^{+1.9}_{-1.8}$\,yr and $t_{\rm break,2} = 24.2^{+3.4}_{-3.8}$\,yr before explosion.

The wind velocity $v_{\rm wind} = 211^{+190}_{-71}$\,km\,s$^{-1}$ is broadly consistent with LBV outflow velocities, though the large uncertainty reflects the degeneracy between wind speed and CSM density profile. Equivalently, the light curve is mainly constraining the density scale of the wind rather than the wind velocity by itself. The supernova parameters ($M_{\rm ej} = 1.2^{+1.3}_{-0.9}\,M_\odot$, $E_{\rm SN} = 2.5^{+1.0}_{-0.5}$~foe) are broadly consistent with a massive progenitor, though the ejecta mass is only weakly constrained by the interaction-dominated light curve. The explosion occurred $76^{+6}_{-7}$ days before first detection, placing the explosion approximately 2.5 months before the earliest observation.

The key physical insight from our analysis is the complex mass-loss history. The steep positive $\alpha_3$ index beyond $\sim$24 years suggests strongly elevated mass loss in the distant past. The intermediate quasi-steady phase ($\alpha_2 \approx 0$, between $\sim$6 and $\sim$24 years before explosion) and the positive final-years slope ($\alpha_1 \approx 1$) suggest a progenitor that underwent elevated mass loss in the distant past, a subsequent quasi-steady phase, and then a decline from several years before explosion toward the final year. The inferred $\dot{M}_0 \sim 0.01\,M_\odot\,\mathrm{yr}^{-1}$ in the final years before explosion is $10^2$--$10^3$ times typical RSG winds, consistent with the high CSM densities inferred from spectroscopy. This broad picture is qualitatively consistent with LBV-like enhanced mass-loss episodes, where dense, relatively slow material can be produced over years to decades~\citeg{Smith2014_review}. Integrating the inferred mass-loss history out to $t_{\rm break,2} \approx 24$\,yr gives a total CSM mass of $1.5^{+1.7}_{-1.1}\,M_\odot$, broadly consistent with the lower bound of $\gtrsim 3\,M_\odot$ from \citet{Fransson2014} given the uncertainty in the unobserved outer CSM. We caution that integrating beyond $t_{\rm break,2}$ gives rapidly diverging estimates because $\alpha_3$ is unconstrained by the data; the total CSM mass is therefore only meaningful within the region probed by our light curve. Within this region, the combined CSM and ejecta mass is a few solar masses, which provides a lower bound on the pre-explosion stellar mass. Our inferred mass-loss rates and wind velocities are broadly consistent with the independent estimates from \citet{Fransson2014}, validating our framework's ability to recover physically meaningful parameters with uncertainties from photometry alone.

%%%%%%%%%%%%%%%%%%%%%%%%%%%%

\subsubsection{SN~2020aeuh}
Type Ia-CSM supernovae represent a rare subclass where thermonuclear explosions interact with pre-existing circumstellar material, providing insights into the progenitor systems of Type Ia supernovae \citep{Silverman2013, Sharma2023}. These events typically show narrow hydrogen emission lines superimposed on Type Ia spectra and often exhibit 91T-like overluminous characteristics, favouring single-degenerate progenitor channels where mass transfer creates the surrounding CSM \citep{Silverman2013}. Recent studies have found that early flux excesses are present in approximately $\sim 20\%$ of SNe~Ia \citep{Deckers2022}, with some events showing evidence for ejecta-CSM interaction at $\sim 10^{14}$~cm as one of several possible explanations~\citep{Sharma2023}.

SN~2020aeuh is classified as a Type Ia-CSM supernova with a distinctive double-peaked bolometric light curve \citep{Weil2021}. The first peak occurs at day $\sim 16$ ($L_{\rm peak} \approx 2.5 \times 10^{43}$~erg~s$^{-1}$), followed by a decline to $\sim 7 \times 10^{42}$~erg~s$^{-1}$, then a secondary plateau/rebrightening phase from days 60--150, before final fading. This double-peaked structure is a smoking-gun indication of Ia-CSM events, distinguishing them from normal thermonuclear supernovae. Analysis by \citet{Tsalapatas2025} suggested that SN~2020aeuh may be a Type Ia interacting with hydrogen- and helium-deficient CSM (Type Ia-CSM-C/O), with CSM mass of $1$--$2\,M_\odot$.

Given the Ia-CSM classification, we model the bolometric light curve compiled by \citet{Tsalapatas2025} with a combined CSM-interaction and radioactive-decay approach with priors suited for Type Ia supernovae. We use the triple power-law wind model of Eq.~\ref{eq:mdot_powerlaw} for the CSM, together with BPL thermonuclear ejecta and radioactive nickel. 
We use the same model as for SN~2010jl with Type-Ia-motivated ejecta and wind priors, listed in Appendix~\ref{app:realfit_tables}: $M_{\rm ej} \in [0.5, 2]\,M_\odot$, $E_{\rm SN} \in [0.3, 3]$~foe, $\dot{M}_0 \in [10^{-5}, 10^{-1}]\,M_\odot\,\mathrm{yr}^{-1}$, $v_{\rm wind} \in [50, 500]$~km~s$^{-1}$, $t_{\rm break,1} \in [0.5, 10]$~yr, and $t_{\rm break,2} \in [10, 100]$~yr. We fix $\kappa = 0.07$~cm$^2$~g$^{-1}$ and $\kappa_\gamma = 0.1$~cm$^2$~g$^{-1}$. An additional nickel fraction prior $f_{\rm Ni} \in [0.1, 1]$ is included to account for radioactive decay (with $M_{\rm Ni} = f_{\rm Ni} \times M_{\rm ej}$; $M_{\rm ej}$ and $v_{\rm ej}$ are derived internally from $M_{\rm ej}$ and $E_{\rm SN}$). The bolometric light curve alone admits a low-nickel solution, but early spectra and Ia template fits favour a radioactively dominated first peak~\citep{Tsalapatas2025}. We therefore also run a spectroscopically motivated variant with a constraint $0.6<M_{\rm Ni}/M_\odot<1.0$.

Without this additional constraint, the light-curve-only fit yields $M_{\rm Ni}=0.27^{+0.12}_{-0.10}\,M_\odot$ and allows the CSM to contribute most of the luminosity near the first peak. This is statistically favoured, but physically uncomfortable as if the first maximum were dominated by interaction, one might expect stronger early interaction signatures and poorer agreement with normal thermonuclear spectral templates. The high-nickel run gives a comparably good fit, with only a modest evidence penalty ($\Delta\ln\mathcal{Z}\simeq -3$) and a maximum likelihood lower by $\simeq1$. In this solution $M_{\rm Ni}=0.66^{+0.06}_{-0.04}\,M_\odot$, $M_{\rm ej}=0.78^{+0.11}_{-0.08}\,M_\odot$, and $E_{\rm SN}=2.10^{+0.55}_{-0.59}$\,foe. The posterior then has the first peak radioactively dominated: at day 16, the nickel contribution is $\sim70\%$ of the bolometric luminosity, while the maximum-likelihood solution has a nickel fraction of $\sim83\%$.

The secondary plateau is produced by CSM interaction when the fast SN ejecta collides with pre-existing material. In the high-nickel wind model, the CSM is characterised by a wind velocity $v_{\rm wind} = 86^{+276}_{-24}$\,km\,s$^{-1}$ and reference mass-loss rate $\dot{M}_0 = 2.2^{+2.2}_{-1.2} \times 10^{-4}$\,$M_\odot\,\mathrm{yr}^{-1}$ one year before explosion. 
The inferred mass-loss history reveals complex episodic behaviour with break times at $t_{\rm break,1} = 7.8^{+1.5}_{-2.2}$\,yr and $t_{\rm break,2} = 31^{+12}_{-13}$~yr before explosion. We find a lower-density CSM than in the unconstrained run. The mass-loss rate declines toward explosion in the final $\sim 8$ years ($\alpha_1 = 0.81^{+0.35}_{-0.43}$ in our lookback-time convention), while the older CSM is less tightly constrained. Integrating the mass-loss history over 0.1--40 yr before explosion gives an effective wind mass of order $M_{\rm csm}\sim0.03\,M_\odot$. We note that the posterior on the mass-loss history is weakly constrained for $t \gtrsim 30$ years before explosion, therefore the credible interval is broad beyond this time. Our inferred mass-loss history suggests a modest but structured episode, possibly a nova eruption or intense mass transfer, with the best-constrained enhancement occurring around several years before explosion and then subsiding during the final years.

This exercise highlights an important degeneracy in bolometric modelling of Ia-CSM events. A low-nickel, early-interaction solution can fit the light curve, but spectral information argues against assigning most of the first peak to CSM interaction. Once this external information is imposed, the first peak is naturally thermonuclear, while the later plateau remains interaction dominated: the CSM contributes $\sim87\%$, $\sim94\%$, and $\sim97\%$ of the model luminosity at 80, 100, and 150 days, respectively. The model therefore supports a picture in which SN~2020aeuh was a thermonuclear explosion with a normal-to-luminous radioactive yield whose later evolution was shaped by interaction with a relatively modest CSM reservoir, consistent with recurrent nova activity, binary mass transfer, or a compact common-envelope-like outflow. We emphasise that the bolometric light curve used here is pseudo-bolometric~\citep{Tsalapatas2025}, and that the radioactive/interaction decomposition should be treated as conditional on the adopted opacity, nickel diffusion prescription, and the spectroscopically motivated nickel-mass prior. In general, the model's ability to
disentangle thermonuclear and interaction contributions at low computational cost demonstrates how our framework can be applied to
interpret hybrid transients where multiple power sources operate simultaneously.

%%%%%%%%%%%%%%%%%%%%%%%%
\subsubsection{SN~2023xgo}
SN~2023xgo is a transitional Type Ibn/Icn supernova at $z = 0.01325$ that exhibited rapid photometric evolution and distinctive spectral characteristics~\citep{Gangopadhyay2025}. Early spectra showed C~{\sc iii}~$\lambda 5696$ emission reminiscent of Type Icn supernovae (interaction with carbon-rich CSM), which later gave way to prominent He~{\sc i} features characteristic of Type Ibn supernovae~\citep{Gangopadhyay2025}. The He~{\sc i} velocities (1800--10000~km~s$^{-1}$) and pseudo-equivalent widths are among the highest in the Ibn/Icn class, suggesting strong interaction. The light curve peaked at $M_r = -17.65 \pm 0.04$ (the faintest in the SN~Ibn sample) with a 5-day rise time and subsequent decline at 0.14~mag~day$^{-1}$ in the optical.

Previous semi-analytical modelling by \citet{Gangopadhyay2025} suggested two distinct CSM-interaction possibilities: a dense compact shell at $\sim 10^{12}$--$10^{13}$~cm with mass-loss rate $10^{-4}$--$10^{-3}$~M$_\odot$~yr$^{-1}$ and $M_{\rm csm} \sim 0.22$~M$_\odot$, versus an extended CSM ($\sim 10^{15}$~cm) with much higher mass-loss rate of $0.1$--$2.7$~M$_\odot$~yr$^{-1}$. The proposed progenitor is an ultra-stripped helium star ($\sim 3$~M$_\odot$) in a close binary system. However, these contrasting CSM profiles would imply significant differences in the progenitor's life before core-collapse.

We fit the observations of SN~2023xgo from \citet{Gangopadhyay2025} (bolometric light curve up to $60$d post-maximum) with a flexible CSM parameterisation, where the CSM is shaped by a time-varying mass-loss history. We compare three scenarios: an eruption-shaped CSM, a power-law density profile on radius combined with a shell, and the variable triple power-law wind of Eq.~\ref{eq:mdot_powerlaw}. The adopted fit uses a smoothed triple power-law wind with priors listed in Appendix~\ref{app:realfit_tables}; in particular, $\kappa = 0.07$~cm$^2$~g$^{-1}$, $M_{\rm expl} \in [0.05, 10]\,M_\odot$, $E_{\rm expl} \in [0.01, 5]$~foe, $\dot{M}_0 \in [10^{-12}, 10^{-2}]\,M_\odot\,\mathrm{yr}^{-1}$, $v_{\rm wind} \in [10, 2000]$~km~s$^{-1}$, $t_{\rm break,1} \in [0.1, 10]$~yr, and $t_{\rm break,2} \in [1, 100]$~yr. The explosion offset prior is $t_{\rm offset} \in [-20, -2]$~d (negative, as data are phased from peak).

The best fit according to the Bayesian evidence is the variable wind, with a log-evidence difference of $\Delta\ln\mathcal{Z} \approx 245$ against the second-best model of a wind + shell. Within this wind parameterisation, the posterior favours $v_{\rm wind} = 996^{+178}_{-96}$\,km\,s$^{-1}$, characteristic of Wolf-Rayet or helium-star winds (typical range $500$--$2000$\,km\,s$^{-1}$)~\citep[e.g.,][]{Smith2014_review}. This should not be read with the same confidence as a direct spectroscopic wind-velocity measurement, but it does disfavour very slow red-supergiant-like winds within the adopted model. The reference mass-loss rate $\dot{M}_0 = 6.6^{+1.6}_{-1.3} \times 10^{-3}\,M_\odot\,\mathrm{yr}^{-1}$ at one year before core collapse is elevated compared to typical WR stars ($10^{-5}$--$10^{-4}$\,M$_\odot$\,yr$^{-1}$), consistent with the enhanced pre-explosion activity inferred for Ibn/Icn progenitors.

The time-variable mass-loss history reveals that both break times are within a few years of explosion ($t_{\rm break,1} = 0.19^{+0.01}_{-0.01}$~yr and $t_{\rm break,2} = 5.6^{+23.3}_{-3.9}$~yr). As $t$ in Eq.~\ref{eq:mdot_powerlaw} is a lookback time before explosion, the negative slopes imply increasing mass loss toward core collapse: $\alpha_1 = -2.44^{+0.07}_{-0.08}$ gives a steep rise in the final months, while $\alpha_2 = -1.15^{+0.16}_{-0.14}$ gives a slower rise through the preceding years. The distant-past slope, $\alpha_3 = -2.28^{+2.79}_{-3.31}$, is essentially unconstrained. This can be seen visually in Fig.~\ref{fig:real_densities}. Integrating the mass-loss history over the region constrained by the lightcurve gives $M_{\rm csm}\approx1.4^{+0.3}_{-0.2}\,M_\odot$, though this should be interpreted as the mass in the effective wind-like profile rather than a directly imaged shell. Overall, our inference suggests enhanced mass loss over the final few years, with the activity intensifying toward the final months before core collapse rather than having already subsided by explosion.

We find a strong preference for the continuous wind model over discrete shell models. This suggests that SN~2023xgo's smooth decline is more naturally explained by sustained interaction with extended CSM rather than the sharper luminosity drop expected when the shock traverses a compact shell. Notably, we find that the entire light curve can be modelled with CSM interaction alone, i.e., no radioactive decay contribution is needed. This is consistent with the identification of narrow lines throughout the spectral series~\citep{Gangopadhyay2025}. We infer an explosion time of $5.2^{+0.01}_{-0.01}$ days before peak luminosity. The fitted ejecta mass ($M_{\rm ej} = 0.36^{+0.07}_{-0.04}\,M_\odot$) and explosion energy ($E_{\rm SN} = 0.084^{+0.053}_{-0.022}$~foe, $v_{\rm ej} \approx 4800$\,km\,s$^{-1}$) point to a low-mass stripped-envelope explosion. This is compatible with the broad ultra-stripped picture proposed by \citet{Gangopadhyay2025}, but the ejecta mass is somewhat above the lowest ejecta masses often quoted for ultra-stripped explosions. We therefore interpret the result more conservatively as evidence for a compact, strongly stripped progenitor in a binary system, rather than as a unique identification of an ultra-stripped supernova channel.

%%%%%%%%%%%%%%%%%%%%%%%%
\subsubsection{SN~2015bn}
SN~2015bn is a nearby ($z = 0.1136$) hydrogen-poor superluminous supernova (SLSN-I) with peak absolute magnitude $M_U \approx -23.1$, also making it one of the most luminous transients ever observed~\citep{Nicholl2016a}. \citet{Nicholl2016a} presented comprehensive multi-wavelength observations from UV to NIR, showing that the light curve and spectra could be explained by a magnetar-powered explosion with spin-down timescale $\sim 20$~days reheating $\sim 10$~M$_\odot$ of oxygen-dominated ejecta. Alternative scenarios include interaction with $\sim 20$~M$_\odot$ of dense, inhomogeneous CSM~\citep{Nicholl2016a}. Notably, the late-time spectrum at $+400$~days closely resembles energetic Type Ic supernovae such as SN~1998bw and SN~2012au, suggesting a connection between superluminous-supernovae and gamma-ray burst progenitors~\citep{Nicholl2016b}, which is plausible for both the magnetar and CSM-interaction models.

For SN~2015bn we fit both the triple power-law wind model (Eq.~\ref{eq:mdot_powerlaw}) and a generic density profile parameterisation. For the latter, we directly parameterise $\rho_{\rm csm}(r)$ as a combination of a background power-law wind and a single Gaussian shell,
\begin{equation}
\rho_{\rm csm}(r) = \rho_{\rm base}\left(\frac{r}{r_0}\right)^{s} + \rho_1\exp\left[-\frac{(r - r_1)^2}{2\,\sigma_1^2}\right],
\label{eq:generic_csm}
\end{equation}
where $\rho_{\rm base}$ and $s$ describe the background wind (with $s = -2$ recovering a steady wind), $r_0 = 10^{14}$~cm is the reference radius, and $\rho_1$, $r_1$, $\sigma_1$ are the peak density, centre radius, and width of the shell. The width is parameterised as a FWHM, $\Delta r_1$, with $\sigma_1 = \Delta r_1 / 2.355$. This generic density profile is used as input to the same Fortran CSM interaction solver, with BPL ejecta. The priors for the generic profile are: $\rho_{\rm base} \in [10^{-16}, 10^{-12}]$~g~cm$^{-3}$ (LU), $s \in [-3, -1]$ (U), $r_1 \in [10^{14}, 10^{16}]$~cm (LU), $\Delta r_1 \in [5\times10^{13}, 5\times10^{15}]$~cm (LU), $\rho_1 \in [10^{-19}, 10^{-11}]$~g~cm$^{-3}$ (LU), $M_{\rm ej} \in [2, 20]\,M_\odot$ (U), $E_{\rm SN} \in [0.5, 10]$~foe (LU), $\epsilon \in [0.1, 0.9]$ (U), and $\kappa = 0.2$~cm$^2$~g$^{-1}$ (fixed). For the alternative wind model we use the same triple-power-law parameterisation with $t_{\rm break,1} \in [0.5, 5]$~yr, $t_{\rm break,2} \in [5, 20]$~yr, $\dot{M}_0 \in [10^{-4}, 1]\,M_\odot\,\mathrm{yr}^{-1}$, $v_{\rm wind} \in [10, 2000]$~km~s$^{-1}$, $M_{\rm ej} \in [2, 20]\,M_\odot$, $E_{\rm SN} \in [1, 50]$~foe, and $\kappa = 0.2$~cm$^2$~g$^{-1}$. Both fits use an explosion offset prior $t_{\rm offset} \in [-150, -10]$~d.

We fitted the bolometric light curve from \citet{Nicholl2016a} (observations from $-79$ to $+392$ days relative to peak). The variable wind model achieves a reasonable fit, but it does so by creating a sharply peaked mass-loss history: $\alpha_1 = 3.35^{+0.37}_{-0.42}$ transitions to $\alpha_2 = -3.87^{+0.11}_{-0.09}$ at $t_{\rm break,1}=3.49^{+0.71}_{-0.62}$\,yr, with $t_{\rm break,2}=16.6^{+2.1}_{-3.2}$\,yr. This profile is effectively using a continuous wind parameterisation to mimic a discrete eruptive enhancement. The corresponding wind velocity is $v_{\rm wind}=1408^{+352}_{-368}$\,km\,s$^{-1}$, the inferred wind mass over the constrained region is $1.24^{+0.91}_{-0.51}\,M_\odot$, and the explosion parameters are energetic ($M_{\rm ej}=15.0^{+3.2}_{-3.9}\,M_\odot$, $E_{\rm SN}=28.1^{+10.0}_{-8.7}$~foe).

The generic density profile provides a more direct description of the same physical picture. Its snapshot of the inferred CSM at the time of the supernova is shown in Fig.~\ref{fig:real_densities}. We infer a shell radius $r_1 = 2.85^{+0.49}_{-0.36} \times 10^{15}$~cm ($\sim 190$~AU) with FWHM $\Delta r_1 = 5.55^{+0.78}_{-0.85}\times10^{14}$~cm, embedded in a background wind with density index $s = -2.10^{+0.63}_{-0.57}$, consistent with a steady wind. The shell density is high, $\rho_1 = 2.2^{+1.2}_{-0.7}\times10^{-12}$\,g\,cm$^{-3}$, and integrating over the fitted density profiles gives a spherical-equivalent CSM mass $M_{\rm CSM}=69.9^{+9.6}_{-8.4}\,M_\odot$. This value should not be read as a literal isotropic shell mass and using the scaling described above, a shell with $f_\Omega f_{\rm fill}=0.1$ would correspond to a physical mass of order $7\,M_\odot$. The broad Gaussian shell may be standing in for clumpy or asymmetric material, or for an extended phase of elevated mass loss rather than a single coherent spherical eruption.

The generic density profile gives more plausible explosion parameters than the variable-wind fit, with $M_{\rm ej}=17.0^{+2.2}_{-3.7}\,M_\odot$ and $E_{\rm SN}=6.9^{+1.6}_{-1.5}$~foe, and an explosion time $100^{+11}_{-9}$ days before peak. The fact that both the triple-wind and generic-shell fits converge on a wind-like background plus a localised density enhancement suggests that the light curve is not simply demanding a smooth steady wind. Instead, SN~2015bn is consistent with an energetic explosion interacting with CSM shaped by eruptive pre-supernova mass loss. This is qualitatively consistent with the alternatives discussed by \citet{Nicholl2016a}. This relatively simple one-shell fit captures the broad bolometric evolution but not the smaller oscillations in SN~2015bn. The framework could fit those structures using the same multi-bump or p-spline density reconstructions used below for SN~2009ip and iPTF14hls, or through an aspherical decomposition, but we do not consider this necessary as these oscillations are consistent with noise at the level of the bolometric data. The resulting progenitor picture is a very massive stripped star that experienced a wind-like mass-loss phase and at least one stronger eruption before explosion. Possible channels include pulsational pair-instability mass loss, a luminous blue variable-like giant eruption in a partially stripped star, or binary/common-envelope mass ejection. In this interpretation the interaction model does not replace magnetar models as a unique explanation for SN~2015bn, but it shows that a wind plus eruptive CSM profile can reproduce the bolometric evolution with physically interpretable structure.

%%%%%%%%%%%%%%%%%%%%%%%
\subsubsection{SN~2009ip}
SN~2009ip is a transient discovered in NGC~7259 in August 2009, initially classified as a supernova before being reclassified as a luminous blue variable (LBV) outburst or ``supernova impostor''~\citep{Smith2010, Foley2011}. Pre-explosion HST imaging from 1999 identified the progenitor as a massive ($M_{\rm ZAMS} \sim 50$--80~M$_\odot$), hot LBV star~\citep{Smith2010, Foley2011}. The source underwent multiple documented outbursts: the initial 2009 event, a subsequent brightening in 2011, and then in 2012 two major events---the 2012a outburst beginning in July (peaking at $M_V \approx-14$~mag) followed by the 2012b re-brightening in September (reaching $M_R \approx -18.5$~mag)~\citep{Mauerhan2013, Fraser2013, Pastorello2013}.

The nature of the 2012 events sparked intense debate. \citet{Mauerhan2013} argued that the broad emission components (FWHM $\sim 8000$~km~s$^{-1}$) and high velocities during 2012a indicated a true
core-collapse supernova during an LBV outburst, with the 2012b brightening arising from ejecta-CSM interaction. In contrast, \citet{Fraser2013} noted the absence of nebular emission from nucleosynthetic
material and an upper limit on $^{56}$Ni mass of $\lesssim 0.02$~M$_\odot$, far below expectations for such a massive progenitor. \citet{Ofek2013} found that the mass-loss rates derived from different diagnostics were inconsistent unless the CSM geometry was highly aspherical. Recent observations appear to have settled this debate, with \citet{Smith2022} arguing that the optical source has faded to 1.2~mag below the quiescent progenitor level, ruling out survival and potentially confirming that the final 2012 event marked as a terminal explosion. Although see~\citet{Pessi2023}, which provides an alternative view on the final fate. 

Given the uncertain nature of the event, but strong evidence for multiple episodes of mass loss, we model the bolometric light curve assembled from \citet{Fraser2013} and \citet{Pastorello2013} with a homologously expanding p-spline CSM density profile combined with $^{56}$Ni radioactive decay. The p-spline is defined by 48 logarithmically-spaced density nodes between $r_{\rm in}$ and $r_{\rm out}$, but is sampled through an initial density, an initial slope, and curvature coefficients. 
The nodes should therefore be interpreted as resolution elements in a smooth density reconstruction, not as 48 independent ``parameters''. The curvature prior regularises the second differences of the log-density profile and suppresses unphysical node-to-node oscillations, while still allowing broad density enhancements where required by the data. This gives the model enough freedom to capture the observed bumps without forcing each feature to be represented by a discrete Gaussian shell (unlike our SN~2015bn example). We use broad priors on the explosion parameters, $M_{\rm ej}\in[0.3,1.4]\,M_\odot$, $E_{\rm SN}\in[0.02,0.4]$~foe, $\epsilon\in[0.2,0.75]$, and $f_{\rm Ni}\in[0,0.04]$, with $\kappa=0.1$~cm$^2$~g$^{-1}$ and  Gaussian priors centred on a maximum-likelihood reconstruction for the spline parameters. The p-spline posterior is sampled locally around the maximum-likelihood solution with \program{emcee}, rather than with nested sampling, because this high-dimensional density reconstruction is more suited to MCMC samplers exploring from a well designed starting point than nested samplers sampling in from a broad prior. There is a risk of overfitting small-scale photometric structure as structure in the density profile With the spline set up. We mitigate this by sampling curvature coefficients rather than independent densities and by imposing a smoothness prior. This combination ensures that unless providing significant statistical benefit, the p-spline with an arbitrary number of nodes collapses to a CSM without any structure, i.e., prioritising a smooth CSM over one with any structure. Regardless we advocate caution when interpreting density profile posteriors from this method and testing which features are robust by reducing the number of spline points and reanalysing the data and testing alternative density profile parameterisations.

The p-spline reconstruction fits the 2012b light curve well and yields a structured CSM with $\log_{10}(r_{\rm in}/{\rm cm})=13.83^{+0.10}_{-0.10}$ and $\log_{10}(r_{\rm out}/{\rm cm})=15.94^{+0.11}_{-0.10}$. The homologous age parameter is $744^{+245}_{-176}$~d, corresponding to material ejected over the few years before the 2012 event for characteristic velocities of hundreds to thousands of km\,s$^{-1}$. The spherical-equivalent CSM mass is $M_{\rm csm}=2.9^{+7.3}_{-1.7}\,M_\odot$. As discussed above, this is a spherical-equivalent density integral; asphericity or clumping would lower the physical mass, which is especially relevant given the geometric arguments of \citet{Ofek2013}. The inferred explosion properties are atypical for a CCSNe. We find $M_{\rm ej}=0.53^{+0.19}_{-0.11}\,M_\odot$, $E_{\rm SN}=0.073^{+0.021}_{-0.016}$~foe, $\epsilon=0.47^{+0.12}_{-0.10}$, and $f_{\rm Ni}=0.0089^{+0.0044}_{-0.0038}$, corresponding to $M_{\rm Ni}\simeq 0.005\,M_\odot$. The nickel mass remains comfortably below the observational upper limit of $\lesssim0.02\,M_\odot$ from \citet{Fraser2013}, so radioactive decay provides only a weak baseline contribution. The luminosity is instead dominated by interaction between low-mass ejecta and dense CSM. This result supports a picture in which SN~2009ip was powered by the collision of relatively low-mass, high-velocity ejecta with CSM produced by the documented pre-2012 activity. The p-spline density profile does not assign one shell to each of the 2009, 2011, and 2012a episodes as directly as the discrete-shell model, but it recovers the same qualitative requirement: the progenitor experienced structured, episodic mass loss shortly before the luminous 2012b event. The inferred low ejecta and nickel masses remain difficult to reconcile with a canonical explosion of a $50$--$80\,M_\odot$ star. They are more naturally interpreted as either a weak terminal explosion, a strongly asymmetric core-collapse event, or an exceptionally violent eruption. This is consistent with the continuing debate over whether the 2012 event was genuinely terminal~\citep{Smith2022, Pessi2023}.

%%%%%%%%%%%%%%%%%%%%%%%
\subsubsection{iPTF14hls}
iPTF14hls is one of the most unusual supernova-like transients discovered by the intermediate Palomar Transient Factory in September 2014~\citep{Arcavi2017}. The event displayed a $\sim 600$-day plateau with at least five distinct brightness fluctuations, while maintaining spectra virtually identical to a normal hydrogen-rich Type II-P supernova~\citep{Arcavi2017}. 
The total radiated energy exceeded $10^{50}$~erg, well above typical Type II-P events, and the bolometric luminosity remained above $10^{42}$~erg~s$^{-1}$ for over 600 rest-frame days, a duration six times longer than canonical supernovae. 
The expansion velocities inferred from spectral lines evolved $5-10$ times slower than expected, and the progenitor location showed a transient in archival plates from 1954, suggesting a possible previous eruption~\citep{Arcavi2017}.

No existing model fully explains iPTF14hls. \citet{Arcavi2017} proposed that the properties could arise from ejection of several tens of solar masses of ejecta a few hundred days before explosion, but offered
no clear mechanism for such ejection. Subsequent studies explored magnetar spin-down~\citep{Dessart2018}, fall-back accretion onto a central remnant~\citep{Wang2018}, pulsational pair-instability supernovae
(PPISN)~\citep{Woosley2018}, as well as variable winds from a massive star~\citep{Moriya2020_iptf}. \citet{Sollerman2019} analysed late-time observations (up to 1000 days) and found evidence for strong circumstellar interaction, with the bumpy light curve possibly arising from ejecta colliding with multiple CSM shells. Such shells could be a natural product of the PPISN channel. \citet{Vigna-Gomez2019} proposed that massive stellar mergers could produce hydrogen-rich PPISN progenitors, as merger products can retain significant hydrogen envelopes up to final explosion, addressing the challenge that single stars massive enough for PPISN typically lose their hydrogen envelopes before core collapse.

We fitted the bolometric light curve from \citet{Arcavi2017} and \citet{Sollerman2019} (spanning $\sim 1000$ days since first observation) with two flexible CSM descriptions. First, we used a 6-shell CSM model superimposed on a power-law wind background, combined with $^{56}$Ni radioactive decay (an extension of the 1-shell model in SN~2015bn). This model is physically readable, since each shell can be associated with a discrete eruption, but it is difficult to sample robustly and can miss some of the smaller undulations, particularly the second and third bumps around $\sim 200$ since first detection. We therefore also fit a 96-node homologously expanding p-spline density profile, using the same local-posterior strategy as for SN~2009ip but with a denser knot set because iPTF14hls has a longer light curve and more structure. As above, the knot values are not independent free shell masses; they are tied together by the p-spline curvature prior, which regularises the density profile while allowing the data to place broad CSM enhancements. The p-spline priors are centred on a maximum-likelihood reconstruction, with $M_{\rm ej}\in[8,20]\,M_\odot$, $E_{\rm SN}\in[0.4,3]$~foe, $\epsilon\in[0.35,0.8]$, and $f_{\rm Ni}\in[0,0.12]$, with the additional constraint $M_{\rm Ni}<2\,M_\odot$.

The 6-shell model captures the gross behaviour of the light curve and favours a massive, interaction-dominated event, but the p-spline reconstruction provides a smoother and more complete description of the observed undulations. The p-spline fit yields $\log_{10}(r_{\rm in}/{\rm cm})=12.80^{+0.19}_{-0.18}$ and $\log_{10}(r_{\rm out}/{\rm cm})=16.23^{+0.03}_{-0.04}$, with a homologous age parameter of $5777^{+1024}_{-922}$~d. The spherical-equivalent CSM mass is $M_{\rm csm}=25.2^{+4.7}_{-3.7}\,M_\odot$. Subject to the same spherical-equivalent caveat described above, this still implies a substantial CSM reservoir and a massive progenitor. The inferred explosion parameters from the p-spline fit are $M_{\rm ej}=13.9^{+1.9}_{-1.8}\,M_\odot$, $E_{\rm SN}=1.08^{+0.13}_{-0.13}$~foe, $\epsilon=0.57^{+0.08}_{-0.07}$, and $n=7.85^{+0.33}_{-0.30}$ for the outer ejecta slope. The radioactive component is non-negligible, with $f_{\rm Ni}=0.066^{+0.031}_{-0.046}$ corresponding to $M_{\rm Ni}=0.91^{+0.48}_{-0.64}\,M_\odot$. This nickel mass is high for ordinary Type II-P supernovae, but lower than the value preferred by the earlier 6-shell fit ($M_{\rm Ni}\simeq1.9\,M_\odot$). 

The physical picture suggested by both the shell and spline analyses is of a massive star embedded in a large, structured CSM reservoir. The 6-shell model makes the episodic interpretation explicit, while the p-spline reconstruction shows that the data require structure over a broad radial range rather than only a small number of isolated shells. The long duration of the light curve, the slow apparent velocity evolution, and the possible 1954 eruption remain naturally compatible with repeated pre-supernova mass ejection over decades. This is qualitatively consistent with pulsational pair-instability or other very massive-star eruption scenarios discussed by \citet{Woosley2018}, \citet{Vigna-Gomez2019}, and \citet{Moriya2020_iptf}, although the light curve alone cannot distinguish between these channels.

\section{Conclusions}\label{sec:discussion}
We have presented a fast, generalised framework for modelling the lightcurves and multi-wavelength diagnostics of interaction-powered transients. The framework numerically integrates the thin-shell equations of motion in \program{Fortran} (with a restrictive \program{JAX}-implementation for inference on GPUs), supports arbitrary CSM density and velocity profiles, and is available through the \program{Redback} inference package. Our main findings are as follows.

\begin{itemize}

\item The framework can reproduce a wide diversity of interaction-powered lightcurves from a common set of equations, spanning simple stellar wind CSM to complex multi-eruption histories. The resulting lightcurves vary by orders of magnitude in peak luminosity and timescale, illustrating both the richness of the observable phenomenology and the interpretational flexibility of the model. The explicit distinction between the one-zone and transport modes also makes clear when the observed luminosity is post-processed from the instantaneous shock power and when photon trapping, diffusion, and post-emergence cooling are being modelled with an explicit radiation field. The transport mode is currently best viewed as a finite static-shell calculation, while the one-zone mode is the inference workhorse for arbitrary CSM histories. This stresses the need for post-fit checks, including comparisons to detailed numerical simulations where possible, to avoid over-interpreting flexible lightcurve fits.

\item A limited comparison against a radiation-hydrodynamical lightcurve further demonstrates the role of different luminosity treatments. The one-zone calculation can reproduce the broad luminosity scale and late-time decline once the CSM becomes optically thin, especially when the time-dependent $\epsilon(t)$ prescription is included. However, the early dark phase and the diffusion-mediated peak require an explicit trapped-radiation reservoir, which is captured only by the transport calculation. The transport mode is nevertheless restricted to static, finite CSM, which may not always be valid. We therefore treat both one-zone and transport models as useful approximations, and advocate testing whether inferred constraints are robust across modelling assumptions or explicitly marginalising over uncertain parameters such as efficiency.

\item The assumed CSM velocity profile has a significant impact on the resulting lightcurve, even when the density structure at the time of explosion is identical. Homologously expanding ejecta-like CSM ($v = r/t$) and wind-like CSM ($v = \mathrm{const}$) can produce markedly different lightcurves at intermediate and late times. This has implications for any inference built on models that assume constant-velocity wind CSM when the true CSM was produced by an eruptive event.

\item Aspherical, two-component CSM can produce double-peaked lightcurves that are qualitatively indistinguishable from two discrete spherical shells under the assumption of spherical symmetry. This underscores the need to consider CSM geometry when interpreting multi-peaked lightcurves.

\item A synthetic inference test on a Gaussian wind model demonstrates that photometric data alone can constrain the effective shape of the mass-loss history, but does not necessarily separate $\dot{M}$ and $v_{\rm wind}$ individually with high confidence, because wind-like bolometric lightcurves mainly constrain the density scale $\dot{M}/v_{\rm wind}$. Independent wind-velocity, shock-velocity, or multi-wavelength constraints can turn this mass-loading measurement into a more direct physical mass-loss-rate estimate.

\item We applied the framework to six diverse interaction-powered transients spanning Type IIn (SN~2010jl), Type Ibn/Icn (SN~2023xgo), Type Ia-CSM (SN~2020aeuh), superluminous (SN~2015bn), LBV-like SN~2009ip, and the extraordinary long-duration event iPTF14hls. SN~2010jl is consistent with LBV-like mass loss over decades, SN~2023xgo with a compact stripped-envelope progenitor and enhanced Wolf-Rayet-like wind, and SN~2020aeuh with a thermonuclear explosion plus a modest CSM reservoir. For SN~2015bn, both a triple-wind model and a generic shell model point toward a wind-like background plus an eruptive density enhancement, although the spherical-equivalent CSM mass is strongly geometry dependent. For SN~2009ip and iPTF14hls, p-spline density reconstructions reproduce the detailed lightcurve structure and favour substantial, structured CSM reservoirs.

\item The framework naturally extends to radio wavelengths via synchrotron emission from the forward shock, and to approximate X-ray diagnostics through thermal bremsstrahlung from the shocked gas. The radio and X-ray examples show that multi-wavelength observables need not trace the optical lightcurve one-to-one: density enhancements can be muted in optical bands by diffusion and temperature evolution, while radio and free-free X-ray emission respond directly to the shock density, velocity, and emission measure. The X-ray calculation includes an explicit shock-power cap, so it should be interpreted as an energy-limited diagnostic rather than a resolved cooling-layer model. These multi-wavelength calculations provide complementary constraints on the CSM density, shock velocity, absorption column, and microphysics that are partly independent of the assumptions entering the optical lightcurve.

\end{itemize}

The framework has a number of limitations that should be kept in mind. An important distinction is between \textit{fit flexibility} and \textit{physical recovery}: the thin-shell equations have sufficient freedom to match a wide variety of lightcurve morphologies, but a good fit does not guarantee that the inferred physical parameters are unique or physically meaningful. This is particularly relevant for high-dimensional models (e.g., multi-shell or spline fits) where strong parameter degeneracies exist, and where additional constraints from spectroscopy, pre-explosion imaging, radio or X-ray observations are needed to break them. The CSM masses quoted for generic shells and spline reconstructions are spherical-equivalent masses. They scale with covering fraction and filling factor, so they should not be interpreted as literal isotropic ejecta masses without independent geometric constraints. The thin-shell approximation and grey transport treatment are also simplifications that may break down in optically thick CSM or when the CSM has complex multi-dimensional structure. The post-emergence cooling phase is also less tightly constrained by the thin-shell dynamics than the interaction phase, because the internal structure of the shocked shell is not resolved. This reinforces the need for more detailed follow-up of inferences with radiative transfer or hydrodynamical simulations as well as inference-capable models relying on other approaches to the problem, such as surrogate modelling or phenomenologically-motivated extensions~\citep[e.g.,][]{Sarin2025, Sarin2026}, and marginalising over uncertain physics instead of forcing narrow posteriors through implicit modelling choices. For the same nominal ejecta, CSM power-law profile, and constant efficiency, our one-zone mode produces similar bolometric light curves to the semi-analytic models of \citet{Chatzopoulos2013}. However the framework produces significantly different light curves in the transport mode and for time-varying efficiencies for the same initial conditions. Similarly, the lightcurves with eruptive mass-loss or more complex mass-loss histories are qualitatively different. In comparison to surrogate models of Type II SNe~\citep{Sarin2025}, we note that the pure effect of CSM interaction is similar across the two approaches, especially when considering the time-varying efficiency (see Fig.~\ref{fig:hydro_comparison}), however the surrogate-model is a more self-consistent calculation of the hydrodynamics, i.e., without the assumption of a thin-shell and inclusion of radioactivity, which is not self-consistently modelled in this work. The trade-off is that the surrogate models is only valid for a small domain, while the present framework is more flexible for arbitrary CSM histories and applicable to a larger set of transients. The blackbody SED assumption used for broadband photometry neglects the rich spectral features of CSM interaction, which require full radiative transfer modelling. X-ray emission, while a sensitive diagnostic of CSM density, involves additional physical complexity (non-equilibrium ionisation, free-free absorption, inverse Compton scattering, and clumping) that is only approximately captured by our current post-processing model. Finally, the high dimensionality of the parameter space for multi-shell and spline models can challenge standard nested samplers, and more efficient sampling strategies or tighter physically motivated priors are necessary for large-scale population studies of lightcurves indicative of rich eruptive histories.

Looking ahead, the arrival of large statistical samples of interacting transients from the Vera C.\ Rubin Observatory and other wide-field surveys will demand exactly the kind of computationally efficient, flexible modelling presented here. Combining photometric constraints from this framework with complementary information from spectroscopy, radio observations, and pre-explosion imaging will be essential for breaking the degeneracies inherent in lightcurve-only inference, and for connecting the inferred mass-loss histories to stellar evolution theory and hydrodynamical simulations of mass loss in massive stars.

%%%%%%%%%%%%%%%%%%%%%%%%%%%%%%%%%%%%%%%%%%%%%%%%%%%%%%%%%%%%%%%%%%%%%%%%%%%%%%%%
\section*{Acknowledgments} 
We thank Claes Fransson, Tatsuya Matsumoto, Sean Brennan, Ilya Mandel, Steve Schulze, Chris Irwin, Kostas Tsalapatas, Anjasha Gangopadhyay, Takashi Moriya, Wynn Jacobson-Galan and Katie Auchettl for helpful discussions. We also thank Takashi Moriya for sharing comparison hydro lightcurves for our benchmarks. We thank Morgan Fraser and Jesper Sollerman for sharing their lightcurves of SN2009ip and iPTF14hls, respectively. NS acknowledges support from the Kavli Foundation. RH acknowledges support from the RIKEN
Special Postdoctoral Researcher Program for junior scientists. The authors acknowledge the use of ChatGPT to assist with plotting scripts and review of the text. The authors take full responsibility for the content.
%%%%%%%%%%%%%%%%%%%%%%%%%%%%%%%%%%%%%%%%%%%%%%%%%%%%%%%%%%%%%%%%%%%%%%%%%%%%%%%%
\section*{Data Availability}
All real data used in this paper are available from referenced published work or on request to the authors. Models are available at \url{https://github.com/nikhil-sarin/redback-csm} and implemented in \program{Redback}~\citep{Sarin2024} publicly available at \url{https://github.com/nikhil-sarin/redback}. Routines to generate simulated data and perform inference are also available via this package. We also utilised {\sc numpy}~\citep{Harris2020} and {\sc matplotlib}~\citep{Hunter2007} for data analysis and plotting.  

\appendix
\section{Real-transient fit priors and posterior summaries}
\label{app:realfit_tables}
Tables~\ref{tab:appendix_wind_fits}--\ref{tab:appendix_pspline_fits} summarise the priors and posterior credible intervals for the adopted real-transient fits shown in Figs.~\ref{fig:real_lcs} and~\ref{fig:real_densities}. We quote posterior medians with 68\% credible intervals. U and LU denote uniform and log-uniform priors, respectively. For the p-spline models, we do not list every curvature coefficient separately; instead we give the shared prior prescription for the coefficient block, since these coefficients regularise the density reconstruction and are not intended to be interpreted as individual physical shell parameters.

\begin{table*}
\centering
\scriptsize
\caption{Priors and posterior summaries for the adopted triple-power-law wind fits.}
\label{tab:appendix_wind_fits}
\begin{tabular}{p{0.12\textwidth}p{0.22\textwidth}p{0.27\textwidth}p{0.28\textwidth}}
\hline
Object & Parameter & Prior & Posterior \\
\hline
SN~2010jl & $t_{\rm offset}$ (d) & U$[0,100]$ & $75.9^{+6.2}_{-7.3}$ \\
 & $t_{\rm break,1}$ (yr) & U$[0.1,10]$ & $6.52^{+1.90}_{-1.79}$ \\
 & $t_{\rm break,2}$ (yr) & U$[10,30]$ & $24.2^{+3.4}_{-3.8}$ \\
 & $\dot{M}_0$ ($M_\odot$ yr$^{-1}$) & LU$[10^{-3},1]$ & $(7.8^{+24.5}_{-5.3})\times10^{-3}$ \\
 & $\alpha_1,\alpha_2,\alpha_3$ & U$[-3,4]$ & $0.98^{+0.60}_{-0.44}$, $-0.05^{+0.09}_{-0.08}$, $2.10^{+0.25}_{-0.19}$ \\
 & $v_{\rm wind}$ (km s$^{-1}$) & LU$[10,1000]$ & $211^{+190}_{-71}$ \\
 & $\delta,n$ & U$[0,2]$, U$[7,14]$ & $0.99^{+0.55}_{-0.55}$, $10.3^{+2.1}_{-1.9}$ \\
 & $M_{\rm ej}$ ($M_\odot$), $E_{\rm SN}$ (foe) & LU$[0.1,50]$, LU$[0.05,20]$ & $1.20^{+1.30}_{-0.87}$, $2.53^{+1.03}_{-0.55}$ \\
 & $\epsilon$, $\kappa$ (cm$^2$ g$^{-1}$) & U$[0.1,0.8]$, fixed 0.34 & $0.58^{+0.12}_{-0.16}$, fixed \\
\hline
SN~2020aeuh & $t_{\rm break,1}$, $t_{\rm break,2}$ (yr) & U$[0.5,10]$, U$[10,100]$ & $7.8^{+1.5}_{-2.2}$, $31^{+12}_{-13}$ \\
 & $\dot{M}_0$ ($M_\odot$ yr$^{-1}$) & LU$[10^{-5},10^{-1}]$ & $(2.2^{+2.2}_{-1.2})\times10^{-4}$ \\
 & $\alpha_1,\alpha_2,\alpha_3$ & U$[-3,3]$ & $0.81^{+0.35}_{-0.43}$, $-0.41^{+2.96}_{-0.36}$, $2.8^{+0.1}_{-4.0}$ \\
 & $v_{\rm wind}$ (km s$^{-1}$) & LU$[50,500]$ & $86^{+276}_{-24}$ \\
 & $\delta,n$ & U$[0,2]$, U$[8,14]$ & $0.93^{+0.65}_{-0.60}$, $12.9^{+0.8}_{-1.0}$ \\
 & $M_{\rm ej}$ ($M_\odot$), $E_{\rm SN}$ (foe) & U$[0.5,2]$, LU$[0.3,3]$ & $0.78^{+0.11}_{-0.08}$, $2.10^{+0.55}_{-0.59}$ \\
 & $\epsilon$, $f_{\rm Ni}$ & U$[0.1,1]$, U$[0.1,1]$ with $0.6<M_{\rm Ni}/M_\odot<1$ & $0.26^{+0.27}_{-0.08}$, $0.88^{+0.08}_{-0.11}$ \\
 & $M_{\rm Ni}$ ($M_\odot$), $\kappa$ (cm$^2$ g$^{-1}$) & derived, fixed 0.07 & $0.66^{+0.06}_{-0.04}$, fixed \\
\hline
SN~2023xgo & $t_{\rm offset}$ (d) & U$[-20,-2]$ & $-5.21^{+0.01}_{-0.01}$ \\
 & $t_{\rm break,1}$, $t_{\rm break,2}$ (yr) & LU$[0.1,10]$, LU$[1,100]$ & $0.190^{+0.010}_{-0.008}$, $5.6^{+23.3}_{-3.9}$ \\
 & $\dot{M}_0$ ($M_\odot$ yr$^{-1}$) & LU$[10^{-12},10^{-2}]$ & $(6.8^{+1.5}_{-1.4})\times10^{-3}$ \\
 & $\alpha_1,\alpha_2,\alpha_3$ & U$[-8,2]$ & $-2.43^{+0.07}_{-0.08}$, $-1.16^{+0.17}_{-0.15}$, $-2.4^{+2.9}_{-3.4}$ \\
 & $v_{\rm wind}$ (km s$^{-1}$), $f_{\rm smooth}$ & LU$[10,2000]$, U$[0.1,1]$ & $997^{+189}_{-95}$, $0.12^{+0.04}_{-0.02}$ \\
 & $\delta,n$ & U$[0.1,2]$, U$[7,12]$ & $0.23^{+0.19}_{-0.09}$, $11.7^{+0.2}_{-0.4}$ \\
 & $M_{\rm expl}$ ($M_\odot$), $E_{\rm expl}$ (foe) & LU$[0.05,10]$, LU$[0.01,5]$ & $0.36^{+0.07}_{-0.04}$, $0.084^{+0.057}_{-0.022}$ \\
 & $\epsilon$, $\kappa$ (cm$^2$ g$^{-1}$) & U$[0.05,0.8]$, fixed 0.07 & $0.53^{+0.18}_{-0.21}$, fixed \\
\hline
\end{tabular}
\end{table*}

\begin{table*}
\centering
\scriptsize
\caption{Priors and posterior summaries for the adopted SN~2015bn generic CSM fit. Radii and widths are sampled in AU in the model interface. The quoted $M_{\rm CSM}$ is a derived spherical-equivalent density integral over the fitted CSM profile.}
\label{tab:appendix_2015bn_fit}
\begin{tabular}{p{0.28\textwidth}p{0.32\textwidth}p{0.30\textwidth}}
\hline
Parameter & Prior & Posterior \\
\hline
$t_{\rm offset}$ (d) & U$[-120,-80]$ & $-99.9^{+11.3}_{-9.4}$ \\
$\rho_{\rm base}$ (g cm$^{-3}$) & LU$[10^{-16},10^{-12}]$ & $(5.5^{+80}_{-5.1})\times10^{-15}$ \\
$s$ & U$[-3,-1]$ & $-2.10^{+0.63}_{-0.57}$ \\
$r_1$ (AU) & LU$[1,10^4]$ & $190^{+33}_{-24}$ \\
$\Delta r_1$ (AU) & LU$[0.1,10^3]$ & $37.1^{+5.2}_{-5.7}$ \\
$\rho_1$ (g cm$^{-3}$) & LU$[10^{-19},10^{-11}]$ & $(2.2^{+1.2}_{-0.7})\times10^{-12}$ \\
$t_{\rm int}$ (d) & LU$[1,10^5]$ & $(1.0^{+3.1}_{-0.8})\times10^4$ \\
$\delta,n$ & U$[0,2]$, U$[8,12]$ & $1.08^{+0.58}_{-0.67}$, $9.80^{+1.33}_{-1.17}$ \\
$M_{\rm ej}$ ($M_\odot$), $E_{\rm SN}$ (foe) & U$[2,20]$, LU$[0.5,10]$ & $17.0^{+2.2}_{-3.7}$, $6.85^{+1.56}_{-1.51}$ \\
$\epsilon$, $\kappa$ (cm$^2$ g$^{-1}$) & U$[0.1,0.9]$, fixed 0.2 & $0.72^{+0.12}_{-0.14}$, fixed \\
$M_{\rm CSM}$ ($M_\odot$) & derived & $69.9^{+9.6}_{-8.4}$ \\
\hline
\end{tabular}
\end{table*}

\begin{table*}
\centering
\scriptsize
\caption{Priors and posterior summaries for the p-spline CSM fits. The p-spline density is sampled through $\log \rho_0$, an initial slope, and second-difference curvature coefficients. The row labelled ``curvature block'' gives the shared prior prescription and the posterior median of the sample-wise median absolute curvature coefficient.}
\label{tab:appendix_pspline_fits}
\begin{tabular}{p{0.12\textwidth}p{0.23\textwidth}p{0.31\textwidth}p{0.25\textwidth}}
\hline
Object & Parameter & Prior & Posterior \\
\hline
SN~2009ip & $t_{\rm offset}$ (d) & U$[-5,5]$ & $-0.29^{+0.21}_{-0.28}$ \\
 & $\log_{10}(r_{\rm in}/{\rm cm})$, $\log_{10}(r_{\rm out}/{\rm cm})$ & U$[13.55,14.05]$, U$[15.75,16.40]$ & $13.8^{+0.10}_{-0.10}$, $15.9^{+0.11}_{-0.10}$ \\
 & $\log_{10}\rho_0$, $\Delta\log_{10}\rho_0$ & U about MLE, U$[-0.5,0.3]$ & $-12.8^{+0.4}_{-0.5}$, $-0.09^{+0.15}_{-0.16}$ \\
 & curvature block & 46 truncated normals about MLE, $\sigma=0.15$ & $0.129^{+0.023}_{-0.022}$ \\
 & $t_{\rm int}$ (d) & LU$[450,1800]$ & $744^{+245}_{-176}$ \\
 & $\delta,n$ & U$[1.2,2.8]$, U$[10,13.2]$ & $1.92^{+0.31}_{-0.34}$, $11.6^{+0.9}_{-0.9}$ \\
 & $M_{\rm ej}$ ($M_\odot$), $E_{\rm SN}$ (foe) & LU$[0.3,1.4]$, LU$[0.02,0.4]$ & $0.54^{+0.19}_{-0.11}$, $0.073^{+0.021}_{-0.016}$ \\
 & $\epsilon$, $f_{\rm Ni}$ & U$[0.2,0.75]$, U$[0,0.04]$ & $0.47^{+0.12}_{-0.10}$, $0.0089^{+0.0044}_{-0.0038}$ \\
 & $M_{\rm Ni}$, $M_{\rm CSM}$ ($M_\odot$) & derived & $0.005^{+0.002}_{-0.002}$, $3.1^{+7.2}_{-1.8}$ \\
\hline
iPTF14hls & $t_{\rm offset}$ (d) & U$[-30,5]$ & $-11.9^{+4.4}_{-6.3}$ \\
 & $\log_{10}(r_{\rm in}/{\rm cm})$, $\log_{10}(r_{\rm out}/{\rm cm})$ & U$[12.55,13.20]$, U$[15.90,16.55]$ & $12.8^{+0.19}_{-0.18}$, $16.2^{+0.03}_{-0.04}$ \\
 & $\log_{10}\rho_0$, $\Delta\log_{10}\rho_0$ & U about MLE, U$[-0.5,0.7]$ & $-11.1^{+0.4}_{-0.4}$, $0.04^{+0.18}_{-0.18}$ \\
 & curvature block & 94 truncated normals about MLE, $\sigma=0.25$ & $0.161^{+0.026}_{-0.025}$ \\
 & $t_{\rm int}$ (d) & LU$[3500,9000]$ & $(5.8^{+1.0}_{-0.9})\times10^3$ \\
 & $\delta,n$ & U$[0,1]$, U$[7,10]$ & $0.16^{+0.30}_{-0.12}$, $7.85^{+0.33}_{-0.30}$ \\
 & $M_{\rm ej}$ ($M_\odot$), $E_{\rm SN}$ (foe) & U$[8,20]$, LU$[0.4,3]$ & $13.9^{+1.9}_{-1.8}$, $1.08^{+0.13}_{-0.13}$ \\
 & $\epsilon$, $f_{\rm Ni}$ & U$[0.35,0.8]$, U$[0,0.12]$ with $M_{\rm Ni}<2M_\odot$ & $0.57^{+0.08}_{-0.07}$, $0.066^{+0.031}_{-0.046}$ \\
 & $M_{\rm Ni}$, $M_{\rm CSM}$ ($M_\odot$) & derived & $0.91^{+0.48}_{-0.64}$, $25.5^{+4.4}_{-3.9}$ \\
\hline
\end{tabular}
\end{table*}

\bibliographystyle{mnras} 
\bibliography{paper}

% Don't change these lines
\bsp	% typesetting comment
\label{lastpage}
\end{document}